\documentclass[aps, prd, showpacs, superscriptaddress, nofootinbib, onecolumn]{revtex4}
\usepackage{amssymb,amsmath,amsfonts,amsbsy,graphicx,microtype,rotating}

\usepackage{amsmath}
\usepackage{graphicx}
\usepackage{bm}
\usepackage[dvips]{color}
\usepackage{amssymb}
\usepackage{amsfonts}
\usepackage{comment}
\usepackage{todonotes}

\def\be{\begin{equation}}
\def\ee{\end{equation}}
\def\ba{\begin{eqnarray}}
\def\ea{\end{eqnarray}}
\def\nn{\nonumber}

\begin{document}

\title{Vacuum and Gravitons of Relic Gravitational Waves,
and Regularization of Spectrum and Energy-Momentum Tensor}

\author{Dong-Gang Wang}
\email{wdgang@mail.ustc.edu.cn}
\affiliation{CAS Key Laboratory for Research in Galaxies and Cosmology, Department of Astronomy, University of Science and Technology of China, Chinese Academy of Sciences, Hefei, Anhui 230026, China}

\author{Yang Zhang}
\email{yzh@ustc.edu.cn}
\affiliation{CAS Key Laboratory for Research in Galaxies and Cosmology, Department of Astronomy, University of Science and Technology of China, Chinese Academy of Sciences, Hefei, Anhui 230026, China}

\author{Jie-Wen Chen}
\email{chjw@mail.ustc.edu.cn}
\affiliation{CAS Key Laboratory for Research in Galaxies and Cosmology, Department of Astronomy, University of Science and Technology of China, Chinese Academy of Sciences, Hefei, Anhui 230026, China}


\begin{abstract}

The spectrum of relic gravitational wave (RGW)
contains high-frequency divergences, which should be removed.
We present a systematic study of the issue,
 based on the exact RGW solution  that covers the five  stages,
from inflation to the  acceleration,
each being a power law expansion.
We show that the present RGW consists of
 vacuum dominating at $f>10^{11}$Hz and graviton dominating at $f<10^{11}$Hz, respectively.
The gravitons are produced by the four cosmic transitions,
mostly by the inflation-reheating one.
We perform adiabatic regularization to remove vacuum divergences in three schemes:
at present, at the end of inflation, and at horizon-exit,
to the 2-nd adiabatic order for the spectrum,
and the 4-th order for energy density and pressure.
In the first scheme a cutoff is needed to remove graviton divergences.
We find that all three schemes yield the spectra of a similar profile,
and the primordial spectrum defined far outside horizon during inflation
is practically unaffected.
We also regularize the gauge-invariant perturbed
inflaton and the scalar curvature perturbation by the last two schemes,
and find that the scalar  spectra, the tensor-scalar ratio,
and the consistency relation remain unchanged.

\end{abstract}

\maketitle

\section{Introduction}

In inflationary cosmology,
relic gravitational wave (RGW) is  generated during inflation
as the traceless-transverse components of metric
perturbations
\cite{Grishchuk,Grishchuk1997,FordParker1977GW,Starobinsky,Rubakov,Fabbri,AbbottWise1984,
    Ford1987,Allen1988,Sahni, Mendes1995,Giovannini,Tashiro,zhangyang05,Zhang06,Morais2014}.
After reheating, radiation, matter and acceleration  stages of the expansion,
it evolves into a stochastic background in the present universe.
Only slightly affected  by  some astrophysical processes
        \cite{Weinberg,WatanabeKomatsu,MiaoZhang,WangZhang,cheng, Schwarz}
during  the evolution,
RGW  carries unique information about the early universe
besides Cosmic Microwave Background (CMB).
Moreover, existing everywhere and all the time
and having a very broad spectrum over $(10^{-18}-10^{11})$Hz,
RGW has been the target for various GW detectors working at different frequency bands,
such as
 LIGO\cite{LIGO}, Virgo\cite{VIRGO},  GEO\cite{GEO}, and KAGRA\cite{KAGRA},
LISA\cite{LISA},
 Pulsar Timing Array (PTA)\cite{Zhaowen2013,Tong},
WMAP\cite{Komatsu,WMAP9Bennett}, Planck\cite{Planck2014}, BICEP2\cite{bicep,BICEPPlanck2015},
and polarized laser beam detectors \cite{Li2003, TongZhangGaussian}.

During inflation, the low-frequency modes of
RGW are stretched outside of horizon
and  remain  constant, $h_k(\tau)=$cons.
The modes in the  band ($ 10^{-18}-10^{-16} $Hz)
reenter the horizon   around $z\sim 1100$
and leave their imprints on CMB.
The  polarization spectrum
 $C_l^{BB}$ in the detection range $l\sim (10-3000)$
is due to the primordial RGW spectrum
\cite{Komatsu,Planck2013,Planck2014, Das2011,Keisler2011,Friedman2009}.
On the other hand, as we shall see,
the high-frequency ($f>10^{11}$Hz) modes  never exit the horizon
and decreases as $h_k(\tau) \propto  1/a(\tau)$.
These correspond to the vacuum part of RGW,
  giving a spectrum $\propto f^2$ and
leading to UV divergences in the auto-correlation function, the energy density and pressure.
Vacuum divergences also occur in any quantum fields in curved spacetime,
such as  inflaton fields and scalar metric perturbations.
To remove the vacuum divergences,
the normal-ordering of field operators in the flat spacetime
will not be proper  in an expanding universe,
since certain finite portion of the vacuum
do have physical effects.
Parker-Fulling's adiabatic regularization with a minimal subtraction rule
\cite{ParkerFulling1974,ParkerFullingHu1974,Bunch1980,AndersonParker1987,ParkerToms,BirrellDavies,Landete:2013lpa}
has been developed to deal with the issue,
and can apply to quantum fields in an expanding Universe, including RGW.
The vacuum  divergences
of quantum fields can be efficiently subtracted to a desired adiabatic order,
while physically relevant part of the vacuum are kept.
The resulting RGW spectrum after regularization
is consequently suppressed in high-frequencies,
which will serve as the target for the high-frequency GW detectors,
{  such as a  polarized, Gaussian laser beam }
        proposed in Refs.  \cite{Li2003, TongZhangGaussian}.
For the low range ($ 10^{-18}-10^{11} $Hz),
the spectrum may  be also possibly modified by regularization.
To investigate  in a precise manner,
the structure of RGW as quantum field in the present accelerating stage
needs to be explored in details.

In literature on adiabatic regularization of quantum fields during inflation,
different schemes and results were put forward
 \cite{Parker2007,Agullo08,Agullo09,Agullo10,Agullo11,Durrer,Finelli2007,
 Marozzi2011,Woodard:2014jba,delRio:2014aua,Bastero-Gil:2013nja},
and there are disagreements on
the regularized primordial spectrum and its spectral indices defined at low-frequencies.
Even doubts arose as to whether adiabatic regularization is proper
for removing  vacuum divergences.
The previous studies considered only the spectrum in the inflation stage,
but not in the   present stage.
Moreover,
these  studies  relied  on the slow-roll approximation during inflation
\cite{LiddleLyth1992,Stewart,LythLiddle2009,KosowskyTurner, LiddleLythbook,Bassett}.
Sometimes inconsistent treatments have been involved.
For instance,  in solving the field equation,
a slow-roll parameter $\epsilon$ was firstly assumed to be a constant,
but then it was allowed to vary in calculating the spectral running index.
Sometimes the spectrum evaluated at horizon-exit
was used in place of the primordial spectrum evaluated at far outside of horizon.
In fact, the two differ drastically in the slope,
the latter is the one actually referred to in CMB observations,
whereas  the former is not.
These shortcomings will
  bring about uncertainties in the resulting spectrum
                      and its regularization.

In this paper, we shall study of the spectrum, energy density and pressure of RGW
in the expanding universe,
and investigate  the issue of removal of UV divergences of the vacuum of RGW
by adiabatic regularization method.
To this end,
we use the exact solution of RGW   that covers
the whole course of   expansion,
from inflation, reheating, radiation, matter, to the present accelerating stage,
each stage being described by
a power-law scalae factor  $a(\tau) \propto \tau^d$ where $d$ is a constant.
Using  the exact spectrum  and the spectral indices $n_t$ and $\alpha_t$
valid at any time and wavenumber,
we show explicitly how the two spectra mentioned above  differ drastically,
and  derive a  relation between $n_t$ and the slow-roll parameter $\epsilon$.
Then we shall explore  the structure of RGW as quantum field in the present stage,
decompose it into the vacuum and gravitons,
and derive the number of gravitons
generated during the cosmic expansions.
We identify that the vacuum dominates for $f>10^{11}$Hz,
gravitons dominate for $f <10^{11}$Hz,
and both have UV divergences to various extent.

Then, we shall apply the formulation of adiabatic regularization
          and the minimal-subtraction rule
to remove the vacuum  divergences of RGW at a generic time.
By explicit calculations we shall show that
the 2-nd adiabatic order is sufficient
for  the spectrum of vacuum containing quadratic and logarithmic divergences,
whereas the 4-th order is needed
for the vacuum energy density and pressure
containing extra quartic  divergences.
To achieve a convergent, present RGW spectrum,
we shall remove UV divergences, from both vacuum and gravitons.
Three schemes of regularization for vacuum divergences  will be presented,
 each at  a different time:
the present time, the ending of inflation, and the horizon-crossing.
For the first scheme,
we perform  adiabatic regularization
for the spectrum, energy density and pressure of the present vacuum,
and remove  the  graviton divergences  by a cutoff.
In the latter two schemes during inflation,
  we shall first regularize the spectrum during inflation,
and let it evolve into the present spectrum,
 according to the evolution equation.
The regularized  RGW spectra from the three schemes are all practically similar,
except a constant factor in the third scheme,
which can be absorbed into the model energy.
Finally,  in parallel to RGW,
both the gauge-invariant perturbed inflaton and the scalar curvature perturbation
have    exact solutions,
and the  regularization during inflation are extended to these fields straightforwardly.
The regularized spectra are unaffected by adiabatic regularization,
  so are the tensor-scalar ratio $r$ and the consistency relation.

The paper is organized as follows.

In Sec. 2, we give the exact solution of RGW,
  the exact power spectrum, the spectral indices,
valid at any wavelength and any time.
The primordial spectrum is examined in details.

In Sec. 3,
we  analyze the structure of RGW  at the present  stage,
and decompose it into   vacuum  and   gravitons.
The number density of gravitons  is given.
The divergent behavior at high frequencies  are analyzed for
the power spectrum, the spectral energy density and pressure.

In Sec. 4,
 we use the  adiabatic regularization
to remove   UV divergences of RGW  vacuum.
The  2-nd adiabatic order regularization is performed on the spectrum,
and  the 4-th order  on the energy density and pressure.
The  formulae are applied to the inflation and accelerating stages.

In Sec.5, three schemes of regularization at different times are presented:
at the present time,
at the end of inflation,  and  at horizon-exit.

In Sec.6,   regularization is extended to the gauge-invariant perturbed inflaton
and the scalar curvature perturbation during inflation in two schemes.

Sec. 7 gives the conclusions and discussions.

Appendix gives technical specifications of
      the exact solution of RGW
and the  joining condition  of the five stages of expansion.
We use the unit with $c=\hbar= 1$ in this paper.

\section{  Relic Gravitational Waves from Inflation to the Present }

For a flat Robertson-Walker spacetime,  the metric is written as
\be \label{metric}
ds^2=a^2(\tau)[d\tau^2-(\delta_{ij}+h_{ij})dx^idx^j],
\ee
which includes  metric perturbations  $h_{ij}$
in  the synchronous gauge   with  $ h_{00}=h_{0i}=0 $.
The tensor perturbation  part of $h_{ij}$
is the traceless and transverse RGW,
and, to linear order of metric  perturbations,
it satisfies the homogeneous wave equation $\square h^{ij}=0$.
In order to reveal the vacuum structure and graviton content of RGW,
in this paper we take RGW as quantum field,
 and expand it  as follows:
\be\label{Fourier}
h_{ij}  ({\bf x},\tau)=\int\frac{d^3k}{(2\pi)^{3/2}}
    \sum_{s={+,\times}} {\mathop \epsilon  \limits^s}_{ij}(k)
    \left[ a^s_{\bf k}   h^s_k(\tau)e^{i\bf{k}\cdot\bf{x}}
    +a^{s\dagger}_{\bf k}h^{s*}_k(\tau)e^{-i\bf{k}\cdot\bf{x}}\right]
       , \,\,\,\, {\bf k}=k\hat{k},
\ee
where two polarization tensors satisfy
\be\label{polariz}
{\mathop \epsilon  \limits^s}_{ij}(k) \delta_{ij}=0,\,\,\,\,
{\mathop \epsilon  \limits^s}_{ij}(k)  k^i=0, \,\,\,\,
{\mathop \epsilon  \limits^s}_{ij}(k) {\mathop \epsilon  \limits^{s'}}_{ij}(k) =\delta_{ss'},
\ee
and $a^s_{\bf k}$ and $a^{s\dagger}_{\bf k}$ are
the annihilation and creation operators of graviton
satisfying  the   canonical commutation relation
\be \label{aadagger}
 \left[a^s_{\bf k}, a^{r\dagger}_{\bf k'}\right]
    =  \delta_{sr}\delta^3({\bf k-k'}).
\ee
For RGW,
the two  polarization modes $h^+_k$ and $h^{\times}_k$
are assumed to be independent and  statistically  equivalent,
so that the superscript $s=+, \times$ can be dropped,
and the wave equation   is
\ba   \label{evolutionh}
h_k^{''}(\tau)+2\frac{a^{'}(\tau)}{a(\tau)}h_k^{'}(\tau)+k^{2}h_k(\tau)=0 .
\ea
Setting
\be \label{hkuk}
h_k(\tau)= A u_k(\tau)/a(\tau),
\ee
 where  $A $ is  a normalization constant,
  the mode   $u_k $ satisfies the wave equation
\be  \label{evolution}
u_k''(\tau)+\left[k^2-\frac{a''(\tau)}{a(\tau)} \right]u_k(\tau)=0.
\ee
For each stage of  cosmic  expansion of Universe, i.e,
inflation, reheating, radiation dominant, matter dominant
and the present accelerating,
the scale factor  is  a power-law  form $a(\tau) \propto \tau^d$
where $d$ is  a constant  \cite{FordParker1977GW,Grishchuk1997,Zhang06},
and the exact solution of Eq.(\ref{evolution})
is a combination of two Hankel functions,
\be
u_k(\tau ) =\sqrt{\frac{\pi}{2}} \sqrt{\frac{\sigma}{2k}}
      \big[ C_2  H^{(1)}_{d-\frac{1}{2}}(\sigma)
  + C_2 H^{(2)}_{d-\frac{1}{2}} (\sigma)\big],
\ee
where $\sigma=k \tau $,
and $C_1$, $C_2$ are coefficients  determined by continuity
of $u_k $, $u_k' $ at the transition of two consecutive stages.
Thus,  we obtain the  analytical solution $h_k(\tau)$
for the whole course of evolution\cite{Zhang06}.
Appendix gives a detailed account of the coefficients
for these five expanding stages
and the joining conditions between the adjoining stages.
Note that cosmic processes, such as neutrino  free-streaming
        \cite{Weinberg,WatanabeKomatsu,MiaoZhang},
QCD  transition,  and $e^+e^-$ annihilation \cite{WangZhang}
   only slightly modify  the amplitude of RGW
and will be  neglected  in  this study.

In particular, for the inflation stage during which RGW is generated,
one has
\be \label{inflation}
a(\tau)=l_0|\tau|^{1+\beta},\,\,\,\,-\infty<\tau\leq \tau_1,
\ee
where two constants $l_0$ and $\beta$ are the parameters of  the model,
$\tau_1$ is the ending time of inflation \cite{Grishchuk1997,Zhang06}.
The expansion  rate is $H=a'/a^2 = -(1+\beta)/l_0 |\tau|^{2+\beta}$.
In the special case of de Sitter,  the inflation index $\beta=-2$,
one has $l_0^{-1}= H$.
Using observational data WMAP \cite{WMAP9Bennett} of the scalar spectral
index $n_s=0.9608\pm 0.0080$,
one can infer from the relation $n_s-1=2\beta+4$
that  $\beta \simeq -2.02$.
For    $\beta\simeq -2$,
the  expansion of Eq.(\ref{inflation})
is quite general, and describes a class of inflation models.
During inflation, Eq.(\ref{evolution}) becomes
\be \label{equ}
u_k''+\left[k^2-\frac{(1+\beta)\beta}{\tau^2} \right]u_k=0,
\ee
and has a  general solution
\be \label{hankel}
u_k(\tau ) = \sqrt{\frac{\pi}{2}}\sqrt{\frac{x}{2k}}
     \big[ a_1  H^{(1)}_{ \beta+ \frac{1}{2} } (x)
          +a_2  H^{(2)}_{\beta + \frac{1}{2}} ( x) \big],
         \,\,\,\,\, \, -\infty  <\tau\leq \tau_1,
\ee
where $x \equiv k|\tau|$,
the coefficients $a_1$ and $a_2 $ are specified by a choice of the initial condition
during inflation.
Note that  $ H^{(1)}_{\beta+ \frac{1}{2}} (x)=  H^{(2)\, *}_{\beta+ \frac{1}{2}} (x)$.
We  take
\be \label{a1a2}
a_1=0,\,\,\, {\rm and} \,\,\, a_2= -ie^{-i\pi \beta/2},
\ee
so that the mode is given by
\be\label{u}
u_k(\tau ) = \sqrt{\frac{\pi}{2}}\sqrt{\frac{x}{2k}}
              a_2 H^{(2)}_{\beta+ \frac{1}{2}} (x),
\ee
which   is the positive-frequency mode
in high frequency limit $k\rightarrow \infty$,
\be \label{uinfl}
u_k \rightarrow \frac{1}{\sqrt{2k}}e^{-ik\tau}.
\ee
The solution was equivalently written in terms of Bessel's functions
\be \label{musol}
u_k= \sqrt{\frac{x}{2k}}\big[ A_1 J_{\frac{1}{2}+\beta}(x)
                 +A_2 J_{-(\frac{1}{2}+\beta)}( x) \big],\,\,\,\,\, \, -\infty<\tau\leq \tau_1,
\ee
with the  coefficients
$A_1=-\frac{i}{\cos \beta\pi} \sqrt{\frac{\pi}{2}} e^{i\pi\beta/2}$
and $ A_2=iA_1e^{-i\pi\beta}$  \cite{Zhang06}.

We work in Heisenberg picture,
in which RGW is  a quantum field evolving in time,
whereas
Fock space vector of quantum state does not change with time.
In addition to the choice of Eq.(\ref{u}),
we assume further that
the   quantum state during inflation is given by the state vector $|0 \rangle$ such that
\be\label{ask}
 a^s_{\bf k}  |0 \rangle =0, \,\,
\ee
for $ s=+,\times$ and  all $  {\bf k} $,
i.e, no gravitons are initially  present,
only the vacuum fluctuations (zero-point energy) of RGW are  present during inflation.
The  coefficient $A$ in Eq.(\ref{hkuk}) can be determined
by the quantum normalization condition,
 which requires  that, during inflation
in each $\bf k$-mode and each polarization of RGW,
there is a zero point energy $\frac{1}{2}\hbar \omega$
in high frequency limit
\be \label{Ak}
A  \equiv \sqrt{32\pi G} =\frac{2}{M_{Pl}},
\ee
where  $M_{Pl}=1/\sqrt{8\pi G}$ is  the Planck mass.
Thus, RGW  during inflationary stage is taken to be
\be  \label{sol}
h_k(\tau)
=\frac{\sqrt{32\pi G}}{l_0|\tau|^{1+\beta}} \sqrt{\frac{\pi}{2}}\sqrt{\frac{x}{2k}}
              \left(-ie^{-i\pi \beta/2}\right) H^{(2)}_{\beta+ \frac{1}{2}} (x)
,\,\,\, \, -\infty<\tau\leq \tau_1.
\ee
Now, the initial condition during the inflation
is fully specified  by  (\ref{ask}) and (\ref{sol}),
which is referred to as Bunch-Davis vacuum state.
 This choice will be tested by cosmological observations,
 such as those via CMB anaisotropies and polarizations.

\begin{figure}
\centering
\includegraphics[width=0.7\linewidth]{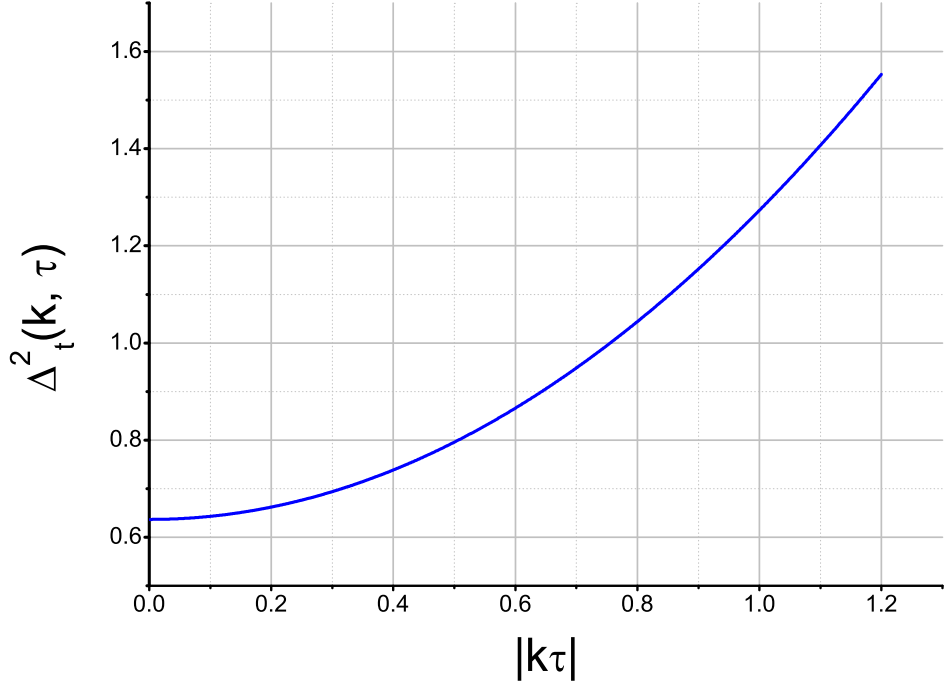}
\caption{The shape of  $\Delta^2_t(k,\tau)$ in Eq.(\ref{gspectrum}) for $\beta=-2$.
It is flat at  $k\tau =0$, but has a steep slope $\propto k^1$
at $k|\tau| = 1$ during inflation.}
 \label{spectrumaroundone}
\end{figure}

\begin{figure}  \label{ntat}
\centering
\includegraphics[width=0.7\linewidth]{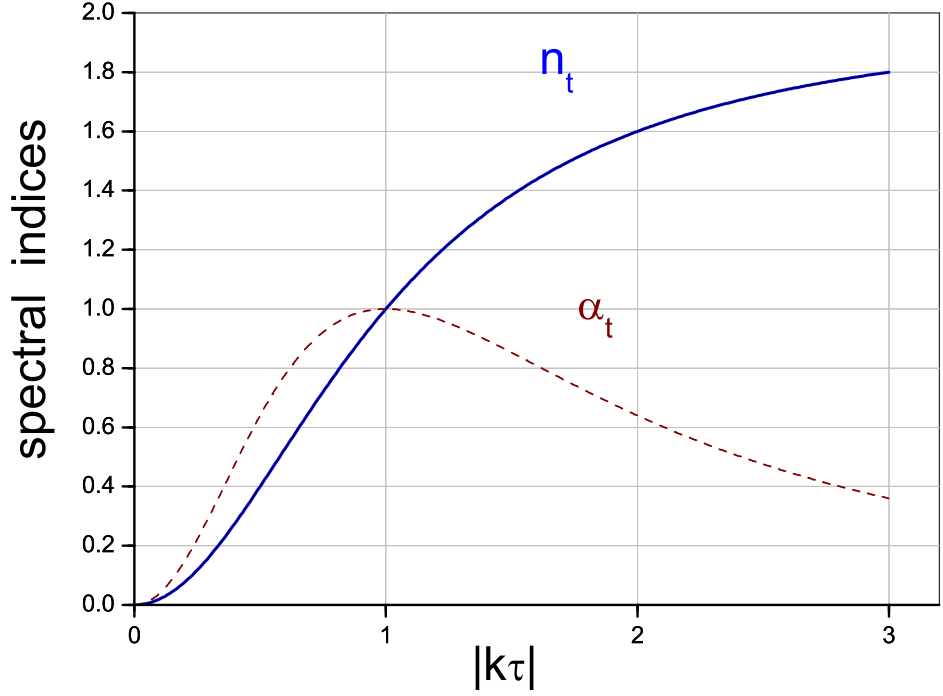}
\caption{The spectral indices  $n_t $ and  $\alpha_t $
defined in (\ref{ntfull})
and (\ref{alphatfull}) as functions of $k|\tau|$ during inflation.
Note that $n_t=\alpha_k=0$ at $k|\tau|=0$, but $n_t=\alpha_k=1$ at $k|\tau|=1$.}
\end{figure}

The auto-correlation function of RGW
is defined as the expectation value of $h_{ij} h^{ij} $,
\ba\label{vevcorr}
\langle0|  h^{ij}(\textbf{x},\tau)h_{ij}(\textbf{x},\tau) |0\rangle
         =   \frac{1}{(2\pi)^3} \int d^3k \,
         (  |{  h_k ^+}|^2 +  |{  h_k  ^\times }|^2) ,
\ea
where Eqs.(\ref {Fourier}) (\ref{polariz})  (\ref{aadagger})  have been used.
The   power spectrum is defined by
\be \label{defspectrum}
\int_0^{\infty}\Delta^2_t(k,\tau)\frac{dk}{k}   \equiv
\langle0|  h^{ij}(\textbf{x},\tau)h_{ij}(\textbf{x},\tau) |0\rangle .
\ee
So one reads off
\be \label{spectrum}
\Delta^2_t(k,\tau)= 2\frac{ k^{3}}{2\pi^2}|h_k(\tau)|^2,
\ee
where the factor 2 is from the polarizations $+, \times$.
In literature  on GW detections,
the characteristic amplitude  $h  (k,\tau)=\sqrt {\Delta^2_t(k,\tau) }$
is often used \cite{Maggiore,Zhang2010}.
The above  definition of   spectrum can be used
for any time $\tau$,  from the inflation to the  accelerating stage.
Notice that,
the  spectrum (\ref{spectrum}) is evolving in the expanding spacetime,
in contrast to the  spectrum in the flat spacetime,
which is independent of time because of the time translation invariance.
Substituting (\ref{sol}) into (\ref{spectrum}) yields
the     exact spectrum  during inflation
\be \label{gspectrum}
\Delta^2_t(k,\tau)= 2\frac{ k^{3}}{2\pi^2a^2}\frac{4}{M_{Pl}^2}|u_k(\tau)|^2
   =\frac{k^{2(\beta+2)} }{2 \pi l_0^2   M_{Pl}^2}
   x^{-(2\beta+1)} H^{(2)}_{\beta+ \frac{1}{2}} (x)   H^{(1)}_{\beta+ \frac{1}{2}} (x)
\ee
Fig \ref{spectrumaroundone} sketches the shape of $\Delta^2_t(k,\tau)$
 as a function of $x=k|\tau|$.
The spectral indices follow from (\ref{gspectrum}) accordingly
\ba \label{ntfull}
n_t(k,\tau)  \equiv\frac{d\ln\Delta_t^2}{d\ln k}
&=&2\beta+4-x   \frac{H^{(2)}_{\beta+ \frac{3}{2}}(x)}{H^{(2)}_{\beta+ \frac{1}{2}}(x)}
-x \frac{H^{(1)}_{\beta+ \frac{3}{2}}(x)}{H^{(1)}_{\beta+ \frac{1}{2}}(x)},
\ea
\ba \label{alphatfull}
\alpha_t (k,\tau) \equiv \frac{d^2\ln \Delta^2_t}{d(\ln k)^2}
&=&(2\beta+1)x    \frac{H^{(2)}_{\beta+ \frac{3}{2}}(x)}{H^{(2)}_{\beta+ \frac{1}{2}}(x)}
-x^2\left[1+ (\frac{H^{(2)}_{\beta+ \frac{3}{2}}(x)}
{H^{(2)}_{\beta+ \frac{1}{2}}(x)} )^2\right]\nn\\
&&+(2\beta+1)x\frac{H^{(1)}_{\beta+ \frac{3}{2}}(x)}{H^{(1)}_{\beta+ \frac{1}{2}}(x)}
-x^2\left[1+  (\frac{H^{(1)}_{\beta+ \frac{3}{2}}(x)}
{H^{(1)}_{\beta+ \frac{1}{2}}(x)} )^2\right].
\ea
For a fixed inflation index $\beta $,
 (\ref{ntfull}) and (\ref{alphatfull}) in long wavelength limit  $x\ll 1$
 give
\be \label{ntk}
n_t  \simeq
 2\beta+4 -\frac{2}{2\beta+3 }x^2 +O(x^3),
\ee
\be \label{alphatk}
\alpha_t
\simeq  -\frac{4}{2\beta+3 } x^2  +O(x^3).
\ee
Fig. \ref{ntat} plots   $n_t$ and $\alpha_t$ as functions of $x$.
These  results hold for the whole class of inflation models
 with $a(\tau) \propto |\tau|^{1+\beta}$.

\begin{figure}
\centering
\includegraphics[width=0.7\linewidth]{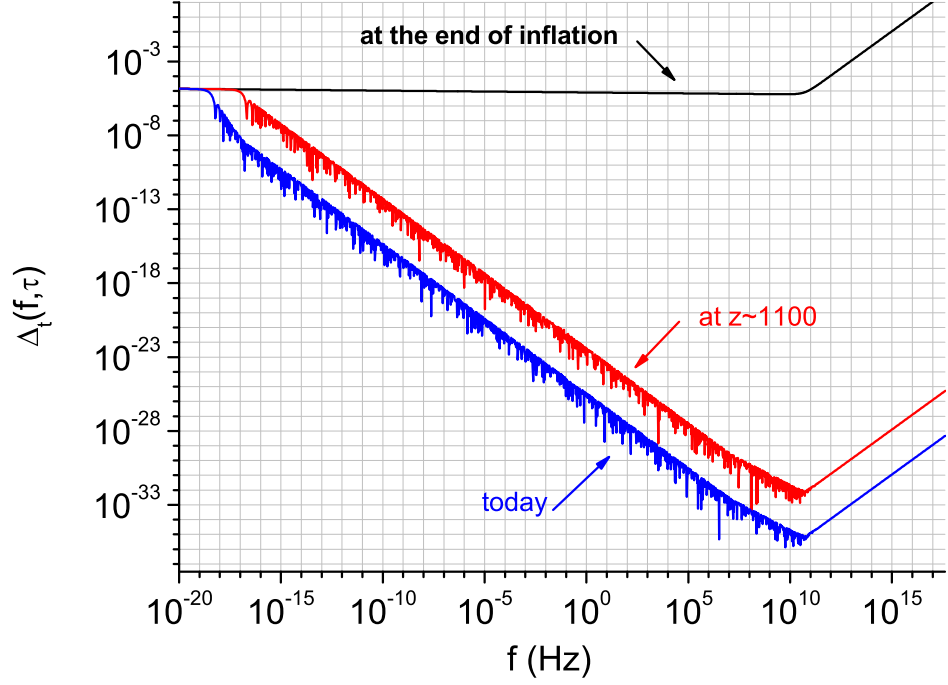}
\caption{ The spectrum $\Delta_t(f,\tau)$ at three different times :
at the end of inflation, at $z\sim1100$, and at present, respectively.
The horizontal axis
is the physical frequency   $f = k/2\pi a(\tau_H)$
at the present time $\tau_H$.
The  parameters $\beta=-2.0125$ and $r=0.12$ are taken for illustration. }
 \label{spectrums}
\end{figure}

The primordial    spectrum
is defined far outside  horizon  $k\ll 1/ |\tau| $ during inflation.
In this long wavelength limit,
the mode of Eq.(\ref{sol}) reduces to
\be  \label{longsol}
h_k(\tau)=\frac{\sqrt{8\pi^2 G}}{l_0}
 \frac{-ie^{i\beta\pi/2}}{\Gamma(\beta+\frac{3}{2})\cos\beta\pi }
           \left(\frac{k}{2}\right)^{\beta+\frac{1}{2}}
      +O( k^{ \beta +\frac{3}{2} } ),
\ee
and the primordial spectrum is given by
\be \label{powerlaw}
\Delta_t^2(k) \equiv \Delta^2_t(k,\tau)  |_{k\ll 1/|\tau|}
    = a_t^2\frac{8}{M_{Pl}^2} \left(\frac{H}{2\pi}\right)^2 k^{2\beta+4}
    \propto k^{2\beta+4} ,
\ee
with
$a_t=   \sqrt{\pi}/   2^{\beta+1}|\Gamma(\beta+\frac{3}{2}) \cos\beta\pi | \simeq 1 $.
Note that (\ref{longsol}) and (\ref{powerlaw})
happen to be independent of $\tau$
as a result of long wavelength limit.
The slope of  (\ref{powerlaw}) depends on the inflation index via $(2\beta+4)$.
In   de Sitter case  $\beta = -2$ and $a_t=1$,
(\ref{powerlaw}) reduces to a flat spectrum \cite{LiddleLythbook}
\be \label{slowroll}
\Delta^2_t(k)
            = \frac{8}{M_{Pl}^2}\left(\frac{H}{2\pi}\right)^2
                       +  O( k^{2}) .
\ee
The primordial spectrum $\Delta_t(k )$  with $\beta=-2.0125$
is shown as the top curve in Fig.\ref{spectrums}.
The spectral indices in limit $k \rightarrow 0$ follow immediately
\be \label{index}
n_t  \equiv\frac{d\ln\Delta_t^2}{d\ln k}| _{k \rightarrow 0}=2\beta+4,
\ee
\be \label{running}
\alpha_t \equiv \frac{d^2 \ln \Delta^2_t}{d(\ln k)^2}|_{k \rightarrow 0}
          = 0.
\ee
This  value of $\alpha_t $
differs from the result of slow-roll approximation \cite{KosowskyTurner,Bassett}.
One can introduce the slow-roll parameter
\be \label{epsilon}
\epsilon \equiv -H'/aH^2 ,
\ee
whose value  is much smaller than $1$.
Solving Eq.(\ref{epsilon}) leads to
\be \label{beta}
\epsilon  =  \frac{\beta+2 }{\beta+1}  .
\ee
Here  $\epsilon$ can be positive or negative,
depending on $\beta  $.
This generalizes  the result $\epsilon >0$
  of a single scalar field model  \cite{LiddleLythbook}.
Plugging Eq.(\ref{beta}) into  Eq.(\ref{index}) yields
the following relation
\be \label{index3}
 n_t =  \frac{-2\epsilon}{1-\epsilon} ,
\ee
which generalizes the result $n_t = -2\epsilon$
of the slow-roll approximation,
but reduces to it when high power terms of $\epsilon$ are dropped.

In regard to   the spectrum and spectral indices,
we would like to point out certain inconsistent treatments in literature.
For instance, sometimes the spectrum and spectral indices around $k|\tau|=1$
were used \cite{LiddleLythbook,KosowskyTurner, Bassett}.
However, here one would have
\be\label{atone}
\Delta^2_t(k)|_{k|\tau| \simeq   1}
            = \frac{8}{M_{Pl}^2}\left(\frac{H}{2\pi}\right)^2
            \left( 2x + O(x-1)^2 \right),
\ee
 $n_t  \simeq 1$, and $\alpha_t  \simeq 1$,
 as seen in Fig.\ref {spectrumaroundone} and Fig.\ref{ntat},
 differing  drastically from (\ref{powerlaw}) (\ref{index}) and (\ref{running}).
(This distinction  applies also to  scalar fields during inflation
as will be addressed   in Sec.6.)
As far as cosmological observations are concerned,
it is incorrect to use  $ k|\tau|= 1$
in place of   $k|\tau| \simeq 0$ for the spectrum and indices.
As is known from analytical calculations of CMB anisotropies and polarization
  \cite{XiaZhang20089, ZhaoZhang,CaiZhang2012},
 the    power spectra  $C_l^{XX}$  located at $l$
are induced  by the $k$-modes of  metric perturbations
in the following manner
\[
C_l^{TT} \propto |h_k(\tau_d)|^2 _{k=l/\tau_H},\,\,\,\,\,
C_l^{EE}, \,  C_l^{BB}  \propto |\dot h_k(\tau_d)|^2 _{k=l/\tau_H} ,
\]
where  $\tau_d$ is the decoupling  time
corresponding to a redshift $z\sim 1100$  when CMB were formed.
Since $C^{XX}_l$  have been observed
in a multipole range   $l\sim (10-3000)$
\cite{Komatsu,Planck2014,Keisler2011,Das2011,Friedman2009},
the relevant   metric perturbation
are those with   \cite{ XiaZhang20089,CaiZhang2012}
\be\label{kmode}
k\sim l/\tau_H  \sim l \sim (10  -3000 ) .
\ee
Among these, the one that entered the horizon exactly at $\tau_d$ is given by
 $k(\tau_d -\tau_m)= 1$, i.e, $k\simeq 26$.
The $k$-modes specified by (\ref{kmode}) stay far outside the horizon
during most part of inflation,
for instance,
they  give
$ k|\tau_1| \sim (6.4\times 10^{-29}- 9.6\times 10^{-26})  \ll 1$
at the end of inflation $\tau_1$.
These modes are just described by
the formulae of  (\ref{longsol}) and (\ref{powerlaw}).
Hence, we conclude that,
it is incorrect  to use $\Delta^2_t(k)$,  $n_t$ and $\alpha_t$ evaluated
at the horizon-crossing to substitute for those at far outside horizon.

The overall amplitude of the primordial spectrum  (\ref{powerlaw})
is essentially determined
by the expansion rate  $H$ of inflation,
which in turn is related to the energy density via $H^2=8\pi G\rho/3$.
In association with observations,
the   spectrum (\ref{powerlaw})
is often rewritten as   \cite{WMAP1}
\be \label{initial}
\Delta_t (k)
=\Delta_{R} \, r^{1/2}(\frac{k}{k_{0}})^{\frac{n_t}{2}
+\frac{1}{4}\alpha_t \ln(\frac{k}{k_0})},
\ee
where $k_{0}$ is a pivot conformal wavenumber
  corresponding to a physical wavenumber
$k_0/a(\tau_H) = 0.002 Mpc^{-1}$,
$\Delta_{R}$ is the curvature perturbation determined by observartions \cite{WMAP9}
$
\Delta_{R}^{2}=(2.464\pm 0.072)\times10^{-9}$,
and $r\equiv \Delta^2_{t}(k_0)/\Delta^2_{R}(k_0)$ is the tensor-scalar ratio,
and $r< 0.12$ by
the joint analysis
of BICEP2/Keck Array and Planck data \cite{MortonsonSeljak2014,BICEPPlanck2015}.

The RGW spectrum at present time $ \tau_H$
follows from  Eq.(\ref{gspectrum})
\be
\Delta^2_t(k,\tau_H)=
    2\frac{ k^{3}}{2\pi^2a^2 (\tau_H)}\frac{4}{M_{Pl}^2}|u_k(\tau_H)|^2 ,
\ee
where   the mode $u_k(\tau_H)$
of the present accelerating stage  has been obtained and is listed in (\ref{upresent}).
The present $\Delta_t(k,\tau_H)$ is plotted as the lowest curve in  Fig.\ref{spectrums}.
Notice  that  $\Delta_t(k,\tau_H)$   is overlapped  with  the primordial one $\Delta_t(k )$
 at the low-frequency end $f< 10^{-18}$Hz,
 both being flat there.
The wavelength of these modes are longer than horizon,
and they remain constant, $h_k\simeq$const,  ever since inflation.
At the  high-frequency end  for $f>10^{11}$Hz,
 the spectrum behaves  as  $\Delta^2 _t(f, \tau_H)\propto f^2$,
as shown in Fig.\ref{spectrums}.
These high-frequency modes  have never exited the horizon since inflation,
so that their amplitude decreases as $h_k  \propto 1/a(\tau)$.
This  high-frequency behavior will cause  the  auto-correlation function
$\langle0|  h^{ij} h_{ij} |0\rangle$ in Eq.(\ref{defspectrum})
to diverge,
an issue to be dressed  in Section 5.

The frequency $f$   at a time $\tau$ is related to the comoving wavenumber $k$
via $f(\tau) = c k/ 2\pi a(\tau)$.
In this paper we adopt the convention
$a(\tau_H)=l_H  \simeq 2.8\times10^{26}$m,
so that the present frequency is related to $k$ via  $f\simeq   1.7\times10^{-19} k$ Hz.
 (see Appendix)

\section{ Decomposition of RGW into  Vacuum and Gravitons }

\begin{figure}
\centering
\includegraphics[width=0.6\linewidth]{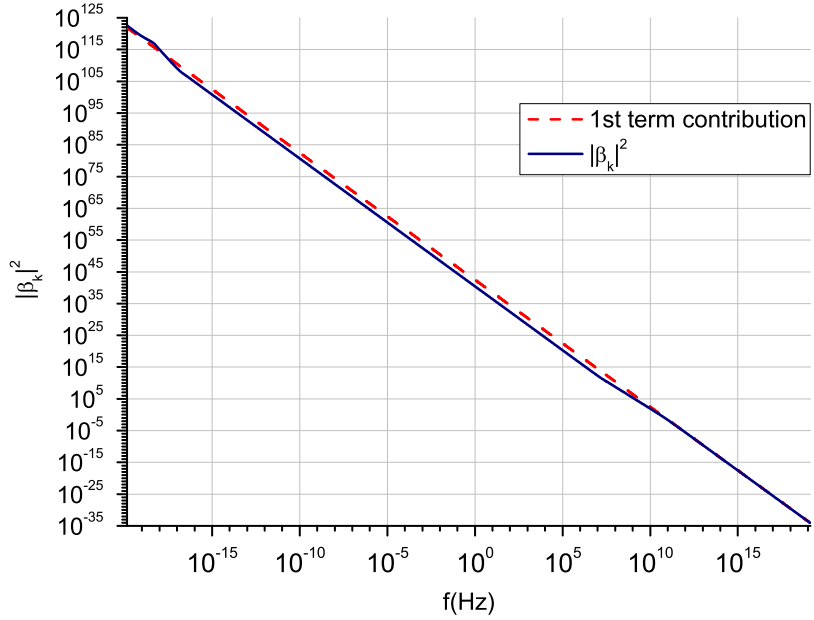}
\caption{The number density $|\beta_k|^2$ of graviton produced in $k$-mode  is shown.
}
 \label{betaksp}
\end{figure}

RGW during inflation has been assumed to be
the vacuum state specified by (\ref{ask}) and (\ref{sol}),
consisting of vacuum fluctuations.
After the subsequent four stages,
RGW has evolved into   the present accelerating  stage
with $a(\tau)=l_H|\tau-\tau_a|^{-\gamma}$,
where $\gamma \simeq 2.1$
fits the model $\Omega_\Lambda \simeq 0.7$ and $\Omega_m =1-\Omega_\Lambda $.
The  analytical   mode    is given by
\be \label{upresent}
u_k(\tau ) = \sqrt{\frac{\pi}{2}}\sqrt{\frac{s}{2k}}
     \big[e^{-i\pi\gamma/2}\beta_k  H^{(1)}_{-\gamma-\frac{1}{2} } (s)
          +e^{i\pi\gamma/2}\alpha_k  H^{(2)}_{-\gamma-\frac{1}{2}} (s) \big],
         \,\,\,\,\, \, \tau_E <\tau\leq \tau_H,
\ee
where $s \equiv k(\tau-\tau_a)$ and
the coefficients $\beta_k, \alpha_k$
are given  by
{ \small
\ba \label{e-1}
e^{-i\pi\gamma/2}\beta_k &=&\Delta_e^{-1}\left\{ \sqrt{\frac{z_E}{s_E}}
\left[d_1H^{(1)}_{\frac{3}{2}}(z_E) +d_2H^{(2)}_{\frac{3}{2}}(z_E)\right]
\left[\frac{1}{2\sqrt{s_E}}H^{(2)}_{-\gamma-\frac{1}{2}}(s_E)
 +\sqrt{s_E} H^{(2)\prime}_{-\gamma-\frac{1}{2}}(s_E)\right] \right. \nn\\
&&\left.-H^{(2)}_{ -\gamma-\frac{1}{2}}(s_E)
\left[\frac{1}{2\sqrt{z_E}}\left(d_1H^{(1)}_{\frac{3}{2}}(z_E)
+d_2H^{(2)}_{\frac{3}{2}}(z_E)\right)\right.\right.\nn\\
&&\left.\left.+\sqrt{z_E}\left(d_1H^{(1)\prime}_{\frac{3}{2}}(z_E)
+d_2H^{(2)\prime}_{\frac{3}{2}}(z_E)\right) \right] \right\},
\ea
\ba \label{e-2}
e^{i\pi\gamma/2}\alpha_k  &=&\Delta_e^{-1}\left\{ \sqrt{\frac{z_E}{s_E}}
\left[d_1H^{(1)}_{\frac{3}{2}}(z_E) +d_2H^{(2)}_{\frac{3}{2}}(z_E)\right]
\left[\frac{1}{2\sqrt{s_E}}H^{(1)}_{-\gamma-\frac{1}{2}}(s_E)+
\sqrt{s_E} H^{(1)\prime}_{-\gamma-\frac{1}{2}}(s_E)\right] \right. \nn\\
&&\left.-H^{(1)}_{-\gamma-\frac{1}{2}}(s_E)
\left[\frac{1}{2\sqrt{z_E}}\left(d_1H^{(1)}_{\frac{3}{2}}(z_E)
+d_2H^{(2)}_{\frac{3}{2}}(z_E)\right)\right.\right.\nn\\
&&\left.\left.+\sqrt{z_E}\left(d_1H^{(1)\prime}_{\frac{3}{2}}(z_E)
+d_2H^{(2)\prime}_{\frac{3}{2}}(z_E)\right) \right] \right\},
\ea
\ba
\Delta_e= \sqrt{s_E}\left[
      H^{(1)}_{-\gamma-\frac{1}{2}} (s_E)  H^{(2)'}_{-\gamma-\frac{1}{2}}(s_E)
     -H^{(1)'}_{-\gamma-\frac{1}{2}} (s_E)H^{(2)}_{-\gamma-\frac{1}{2}} (s_E)\right],
\ea
}
where  $s_E=k(\tau_E-\tau_a)$ and $z_E=k(\tau_E-\tau_m)$,
and  $d_1$ and $d_2$ are the coefficients for the precedent matter  stage.
(See   Appendix for details.)
In high-frequency limit $k \rightarrow \infty$,
$\beta_k , \alpha _k$  have the following asymptotic expressions:
\ba \label{e1}
 \beta _k &=&\left(\frac{\beta(\beta+1)}{4x_1^2}-\frac{\beta_s(\beta_s+1)}{4t_1^2}\right)
e^{i(x_1+t_1-t_s+y_s-y_2+z_2-z_E+s_E)-i\pi\beta }\nonumber\\
&&+\frac{\beta_s(\beta_s+1)}{4t_s^2}e^{i(x_1-t_1+t_s+y_s-y_2+z_2-z_E+s_E)+i\pi\beta }\nonumber\\
&&-\frac{1}{2z_2^2}e^{i(x_1-t_1+t_s-y_s+y_2+z_2-z_E+s_E)+i\pi\beta }\nonumber\\
&&+\left(\frac{1}{2z_E^2}-\frac{\gamma(\gamma+1)}{4s_E^2}\right)
e^{i(x_1-t_1+t_s-y_s+y_2-z_2+z_E+s_E)+i\pi\beta }+\mathcal{O}(k^{-3})
\ea
\ba \label{e2}
 \alpha_k  &=&e^{-i(x_1-t_1+t_s-y_s+y_2-z_2+z_E-s_E)+i\pi\beta }
\left(1-i\frac{\beta(\beta+1)}{2x_1}+i\frac{\beta_s(\beta_s+1)}{2t_1}\right.\nonumber\\
&&\left.-i\frac{\beta_s(\beta_s+1)}{2t_s}+i\frac{1}{z_2}-i\frac{1}{z_E}
+i\frac{\gamma(\gamma+1)}{2s_E}
-\frac{\beta^2(\beta+1)^2}{8x_1^2}-\frac{\beta_s^2(\beta_s+1)^2}{8t_1^2}\right.\nonumber\\
&&\left.-\frac{\beta_s^2(\beta_s+1)^2}{8t_s^2}
-\frac{1}{2 z_2^2}-\frac{1}{2 z_E^2}-\frac{\gamma^2(\gamma+1)^2}{s_E^2}+
\frac{\beta(\beta+1)\beta_s(\beta_s+1)}{4x_1t_1}\right.\nonumber\\
&&\left.-\frac{\beta(\beta+1)\beta_s(\beta_s+1)}{4x_1t_s}
+\frac{\beta(\beta+1)}{2x_1z_2}-\frac{\beta(\beta+1)}{2x_1z_E}+
\frac{\beta(\beta+1)\gamma(\gamma+1)}{4x_1s_E}\right.\nonumber\\
&&\left.+\frac{\beta_s^2(\beta_s+1)^2}{4t_1t_s}
-\frac{\beta_s(\beta_s+1)}{2t_1z_2}+\frac{\beta_s(\beta_s+1)}{2t_1z_E}
-\frac{\beta_s(\beta_s+1)\gamma(\gamma+1)}{4t_1s_E}\right.\nonumber\\
&&\left.+\frac{\beta_s(\beta_s+1)}{2t_s z_2}-\frac{\beta_s(\beta_s+1)}{2t_s z_E}
 +\frac{\beta_s(\beta_s+1)\gamma(\gamma+1)}{4t_s s_E}+\frac{1}{z_2z_E}
 -\frac{\gamma(\gamma+1)}{2z_2s_E}\right.\nonumber\\
&&\left.+\frac{\gamma(\gamma+1)}{2z_Es_E}
\right) +\mathcal{O}(k^{-3}).
\ea
where $x_1, t_1, t_s, y_s, y_2, ..., s_E$ are
 the time instances of transitions multiplied by the wavenumber (see Appendix).
Analogous to Eq.(\ref{u}) for inflation,
the  vacuum mode during the present stage
is   chosen as
\be \label{uvacpres}
v_{k}(\tau )= \sqrt{\frac{\pi}{2}}\sqrt{\frac{s}{2k}}
          e^{i\pi \gamma/2} H^{(2)}_{-\gamma-\frac{1}{2}}(s),
                \,\,\,\,\, \, \tau_E <\tau\leq \tau_H,
\ee
so that
$v_{k}(\tau ) \rightarrow \frac{1}{\sqrt{2k}} e^{-ik(\tau-\tau_a)} $
as  $k\rightarrow\infty$.
Thus, in terms of $v_k(\tau)$, Eq.(\ref{upresent}) is written as
\be \label{upresent2}
u_k(\tau ) = \alpha_k v_k(\tau) +\beta_k  v_k^*(\tau) ,
\ee
and $\alpha_k$ and $\beta_k$ are the Bogolyubov coefficients,
satisfying the relation
\be \label{bogolyubov}
|\alpha_k|^2-|\beta_k|^2=1,
\ee
resulting from the commutation relation  (\ref{aadagger}).
Starting from vacuum fluctuations described by
the positive-frequency mode (\ref{u}) during inflation,
 RGW has  evolved into a mixture of the positive and negative frequency modes
as in Eq.(\ref{upresent2}) for the present  stage.
From   the field operator $h_{ij}$ in Eq.(\ref{Fourier}),
one sees that the operator for  each  $\bf k$ is proportional to
\[
a_{\bf k}u_k +a_{\bf k}^\dagger  u_k^*
              = A_{\bf k} v_k  +A^\dagger_{\bf k } v_k^*,
\]
where
\[
 A_{\bf k} \equiv \alpha_k  a_{\bf k}  + \beta^*_k a^\dagger_{\bf k}
\]
is  interpreted as the annihilation operator of gravitons of $\bf k$
for the present   stage.
Thus,  the number density of gravitons in the present  stage is
\be
N_{\bf k } = \langle  0|  A_{\bf k}^\dagger   A_{\bf k} |0\rangle  = |\beta_k|^2.
\ee
This result is an application
of the theory of particle production in expanding Universe,
developed by Parker \cite{Parker1966}.
As a function of $k$,  $|\beta_k|^2 $
is shown  in Fig.\ref{betaksp},
and over the frequency range $f\ge 10^{-18}$Hz,
one has $|\beta_k|^2 \propto k^{-4}$ approximately,    as shown  in Eq.(\ref{e1}).
Moreover, as   $k\rightarrow\infty$, $\alpha_k\sim1$ and $\beta_k \propto k^{-2}$,
so that $u_k $ of (\ref{upresent2}) is dominated by
the positive frequency mode $ v_k $  in high frequency limit.
This confirms the adiabatic theorem  \cite{Parker1966,Parker1969,ParkerFulling1974},
 i.e,
  high frequency modes  are essentially unaffected by
a slow expansion of the spacetime.
Our detailed calculations show that the modes with $f>10^{11}$Hz
never exit the horizon from  inflation up to the present  stage.

  $\beta_k$ in Eq.(\ref{e1}) contains terms
such as $\frac{\beta(\beta+1)}{4x_1^2}  \propto a''/a$.
By the Friedmann equation
${a''/a}   =\frac{4 \pi G}{3} a^2 T^{\mu}\, _{\mu}$,
it is revealing to express these in  terms of the trace $T^ \mu\,_\mu$
of the energy momentum tensor that drives the cosmic expansion.
One has
\ba \label{betak}
\beta_k&=&\frac{a(\tau_1)^2\pi G}{3k^2}
\left[T^{\mu}\, _{\mu}(\tau_1^-)-T^{\mu}\, _{\mu}(\tau_1^+)\right]
e^{i(x_1+t_1-t_s+y_s-y_2+z_2-z_E+s_E)-i\pi\beta}\nonumber\\
&&+\frac{a(\tau_s)^2\pi G}{3k^2}
\left[T^{\mu}\, _{\mu}(\tau_s^-)-T^{\mu}\, _{\mu}(\tau_s^+)\right]
e^{i(x_1-t_1+t_s+y_s-y_2+z_2-z_E+s_E)+i\pi\beta}\nonumber\\
&&+\frac{a(\tau_2)^2\pi G}{3k^2}
\left[T^{\mu}\, _{\mu}(\tau_2^-)-T^{\mu}\, _{\mu}(\tau_2^+)\right]
e^{i(x_1-t_1+t_s-y_s+y_2+z_2-z_E+s_E)+i\pi\beta}\nonumber\\
&&+\frac{a(\tau_E)^2\pi G}{3k^2}
\left[T^{\mu}\, _{\mu}(\tau_E^-)-T^{\mu}\, _{\mu}(\tau_E^+)\right]
e^{i(x_1-t_1+t_s-y_s+y_2-z_2+z_E+s_E)+i\pi\beta}+\mathcal{O}(k^{-3})\nn\\
\ea
where $T^{\mu}\, _{\mu}(\tau_1^-)$ is evaluated at the end of inflation,
$T^{\mu}\, _{\mu}(\tau_1^+)$  at the beginning of reheating,
etc.
Eq.(\ref{betak}) tells   that the graviton production
 is due to
the discontinuities  at the transitions
of the trace of the energy momentum tensor
that drives the expansion.
In our model,   the pressure $p$ is not continuous at the transition points.
Furthermore, among the four terms in (\ref{betak}),
the first term $\propto  1/(k\tau_1)^2$  by the inflation-reheating transition
gives  the greatest contribution,
other three terms give some modifications.
This analytically confirms
the conclusion that particle creation at the early stages
is of great significance \cite{Parker1968}.
Our computation shows that
the full  $|\beta_k|^2$ computed from (\ref  {e-1})
is lower than the square of the  first term
by two orders in magnitude  within the   range $ (10^{-17} -10^7 )$Hz.

It is interesting to compare our result with
the well known results of production of scalar particles in RW spacetimes.
For a scalar massless field conformally  coupled with the curvature,
there is no particle production of the scalar field\cite{Parker1966,Parker1969,ParkerToms}.
This conclusion holds before regularization
where one has classically    $T^{\mu}\, _{\mu}=0$,
i.e,  the trace of energy momentum tensor
of the  scalar field is vanishing,
as well as after regularization
whereby  the trace anomaly $\langle 0| T^{\mu}\, _{\mu}|0\rangle_{phys} \ne 0$
 appears \cite{ParkerToms,BirrellDavies, Ford1985}.
For cases of non-conformal coupling,
 in general,
the trace $T^{\mu}\, _{\mu}\ne 0$ and,
there are particle productions of the scalar field.
However,
in our expression (\ref{betak}),
 $T^{\mu}\, _{\mu}$ is that of the background
 matter content that drives the expansion and
 may not be the scalar field in Refs.\cite{ParkerToms,BirrellDavies, Ford1985}.

\begin{figure}
\centering
\includegraphics[width=0.7\linewidth]{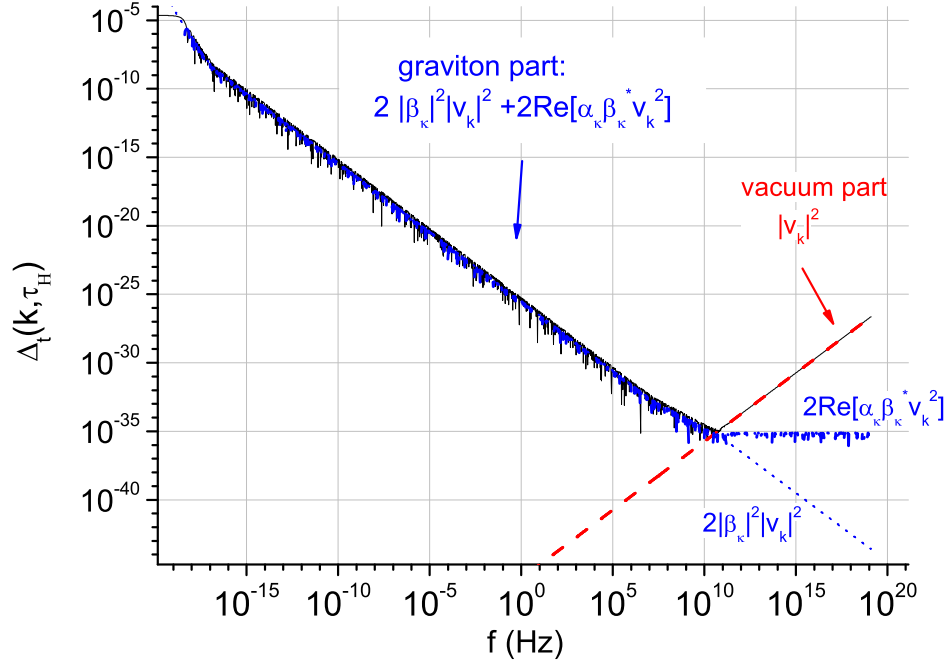}
\caption{ The present spectrum
consists of the vacuum and graviton parts.
At $f> 10^{11}$Hz
the vacuum is dominant
and contains both  quadratic and log  divergences $\Delta^2_t(k) \propto k^2, k^0$,
and the gravitons gives only log  divergence $\Delta^2_t(k) \propto k^0$.}
 \label{dspectrum}
\end{figure}

Now we  analyze the present power spectrum
\be \label{unregulspectrum}
\Delta_t^2(k,\tau_H) = A ^2\frac{k^3}{\pi^2 a^2(\tau_H)}   |u_k(\tau_H)|^2 ,
\ee
in terms of vacuum and gravitons.
By   Eq.(\ref{upresent2})  and the relation in (\ref{bogolyubov}),
one has the following decomposition
\ba  \label{uuvac}
|u_k(\tau )|^2   =  |v_k(\tau)|^2
              + 2Re[\alpha_k \beta_k^* v_k(\tau)^2]
              + 2|\beta_k|^2  |v_k(\tau)|^2,
\ea
where  $|v_{k}|^2 $  is the   vacuum term given by (\ref{uvacpres}),
and the last  two terms containing $\beta_k$ are due to  the  gravitons.
In high frequency limit,
  the vacuum term behaves as
$|v_k |^2   \propto k^{-1}, k^{-3}$,
the  cross  term   as
$  Re[\alpha_k \beta_k^* v_k ^2]\propto k^{-3}$,
and
$   |\beta_k|^2  |v_k |^2 \propto k^{-5}$,
so that the spectrum contains the following quadratic and logarithmic divergences
\be\label{uvspectr}
\Delta^2_t(k,\tau_H)\propto k^2, k^0, \,\, {\rm for}\,\, f> 10^{11}{\rm Hz} ,
\ee
coming from the vacuum  and  cross terms,
which will be removed in later sections.
In the   range $f<  10^{11}$Hz,
$\Delta^2_t(k,\tau_H)$ is dominated by the graviton terms
 $Re[\alpha_k \beta_k^* (v_k )^2]$ + $|\beta_k|^2 |v_k|^2 $,
both having the same   profile,
except that $|\beta_k|^2 |v_k|^2 $  is   smooth,
whereas $Re[\alpha_k \beta_k^* (v_k )^2]$ has extra quick oscillations,
caused by the interference of waves between vacuum $\alpha_k v_k$
and gravitons $\beta_k v_k^*$.
The slope of overall profile of $\Delta^2 _t(k,\tau_H) $ is
$\propto k^{-2+(2\beta+4)}$ in $(10^{-18}-10^{7})$Hz,
and $ \propto k^{-1.5+(\beta+2)}$ in $(10^{7}-10^{11})$Hz.
These features  are  illustrated in Fig.\ref{dspectrum}.

\begin{figure}
\centering
\includegraphics[width=0.6\linewidth]{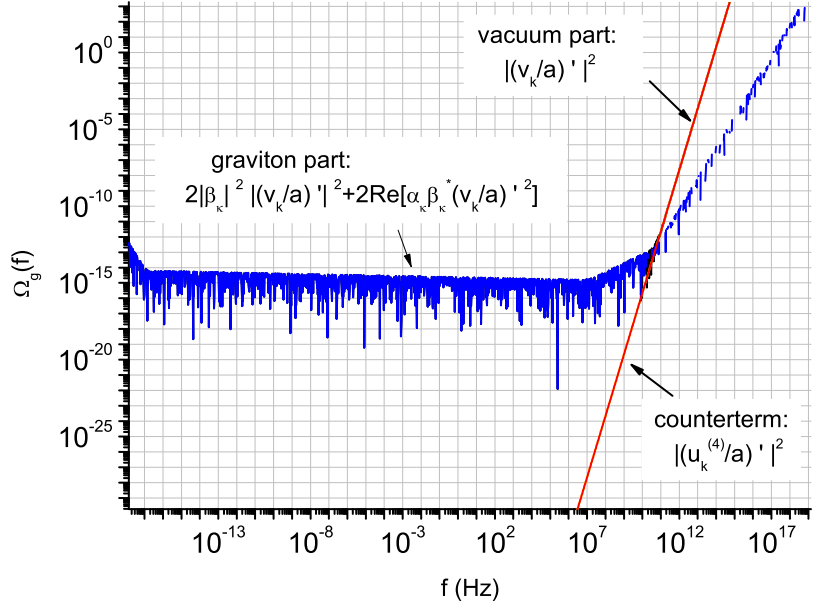}
\caption{    $\Omega_g (f) = \rho_k(\tau_H)/\rho_c$ at present
 consists of the vacuum and graviton parts,
where   $\rho_c$ is the critical density.
       $\beta=-2.0125$ and $r=0.12$ are taken.}
     \label{energy2}
\end{figure}

\begin{figure}
\centering
\includegraphics[width=0.6\linewidth]{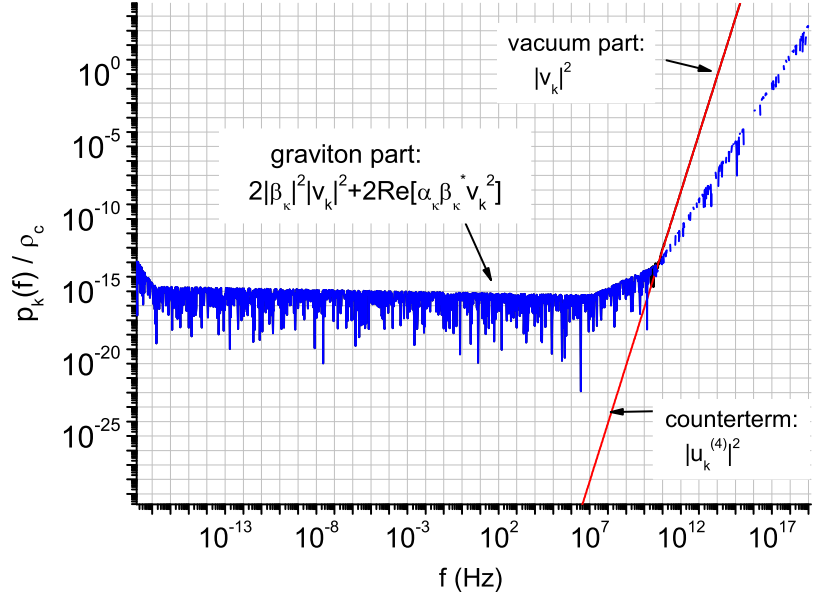}
\caption{ Spectral pressure $p_k(\tau_H )/\rho_c$ at present
consists of the vacuum and graviton parts.
$\beta=-2.0125$ and $r=0.12$. }
 \label{pressure2}
\end{figure}

The energy momentum tensor of RGW can be also decomposed into  vacuum and gravitons.
As long as the wavelengths  are  shorter than the horizon,
i.e, $f\ge  10^{-18}$Hz,
the    energy-momentum tensor of RGW is well defined and given by
                  \cite{Isaacson1968,BrillHartle,SuZhang}
\be\label{tmunu}
t_{\mu \nu }=\frac{1}{32\pi G}\langle 0|h^{ij}_{\ , \, \mu }h_{ij,\, \nu }|0\rangle ,
\ee
 the energy density
\be\label{rho}
\rho_{gw}=t^0\, _0 =\frac{1}{32\pi G a^2}  \langle 0|h'_{ij}   h'\, ^{ij} |0\rangle ,
\ee
and the pressure
\be
p_{gw} = -\frac{1}{3}t^i\, _i.
\ee
Substituting (\ref{Fourier}),   (\ref{polariz}) and  (\ref{aadagger})
into (\ref{rho})  yields
\be \label{energy}
\rho_{gw}   = \frac{1}{32\pi G a^2}\int \frac{d^3 k}{(2\pi)^3}  \, 2 |h_{k}'(\tau)|^2
=\int^{\infty}_0 \, \rho_k(\tau)\frac{d k}{k},
\ee
where the spectral energy density
\be  \label{rhok}
 \rho_k(\tau) = 2\frac{k^3}{2\pi^2a^2} \left| (\frac{u_{k}}{a} )'\right|^2
\ee
with
\be\label{u'}
 \left| ( \frac{u_k }{a} ) ' \right|^2
 =   \frac{|u_k'|^2}{a^2}   + (\frac{a'}{a^2})^2 |u_k|^2
   -\frac{ a'}{a^3} ( u_k^* u_k' + u_k u_k'^* ).
\ee
The formula (\ref{rhok})   holds at any time $\tau$.
Note that, in high-frequencies,
$\rho_k$ is dominated by  the first term  of (\ref{u'})
and can be approximated  by
\be \label{rhohk}
\rho_k (\tau ) \simeq \frac{k^3 }{\pi^2 a^4} \left|    u_{k} '   \right|^2
               \simeq  \frac{k^5 }{\pi^2 a^4} \left|    u_{k}    \right|^2
               =\frac{1}{32\pi G a^2} k^2 \Delta^2_t(k,  \tau ),
\ee
which has been often used in literature \cite{AllenRomaro1999, Grishchuk1997,Zhang2010}.
But for regularization later,
one should  use the full expression (\ref{u'}).
Similar to  (\ref{uuvac}), one has  the vacuum-graviton decomposition
\be \label{u'2}
\left| ( \frac{u_k}{a} ) ' \right|^2
 =   | ( \frac{v_k }{a} ) '  |^2
  +   2Re[ \alpha_k \beta_k^*   ( \frac{v_k }{a} ) ^{\prime\,  2}]
       + 2|\beta_k|^2   | ( \frac{v_k }{a} ) '  |^2   .
\ee
At  $f>10^{11}$Hz,
the  vacuum  term $|\left( \frac{v_k }{a}\right) '|^2 \propto k^1,k^{-1},k^{-3}$,
the cross term gives
\be\label{rhogravitona}
 Re [\alpha_k \beta_k^*   ( \frac{v_k }{a}  ) ^{\prime\,  2}
         ] \propto k^{-1}, k^{-3},
\ee
and
\be \label{rhogravitonb}
|\beta_k|^2   | ( \frac{v_k }{a} ) '  |^2
          \propto k^{-3}  ,
\ee
so that $\rho_k $ contains quartic, quadratic, and logarithmic divergences
\be\label{rhodiv}
\rho_k \propto k^4, k^2, k^0 ,
\ee
 in the  integration (\ref{energy}).
At  $f<  10^{11}$Hz,  the graviton terms
  $Re[ \alpha_k \beta_k^*   ( \frac{v_k }{a} ) ^{\prime\,  2}]$
 +$|\beta_k|^2   | ( \frac{v_k }{a} ) '  |^2   $ dominate,
giving $\rho_{k} \propto k^{2\beta+ 4}$ in $  (10^{-18}-10^{7})$Hz and
$ \rho_{k}  \propto k^{0.5+(\beta+2)} $    in    $  (10^{7}-10^{11})$Hz.
These are illustrated in Fig. \ref{energy2}.
Similarly,
the pressure  is \cite{SuZhang}
\be \label{pre}
p_{gw} =  \frac{1}{96 a^2\pi G}\int \frac{d^3k}{(2\pi)^3}2k^2|h_k|^2
=\int^{\infty}_0 p_k(\tau)\frac{dk}{k} ,
\ee
where
\be \label{pressk}
p_k(\tau) =\frac{k^5}{3\pi^2a^4}|u_k(\tau)|^2
\ee
is  the   spectral pressure.
By (\ref{uuvac}),  it also has the decomposition
\be \label{p}
p_k(\tau)   = \frac{k^5}{3\pi^2a^4}\left[ |v_{k}|^2
    + 2Re( \alpha_k \beta_k^*  v_k ^2 )   + 2|\beta_k|^2  |v_k   |^2  \right].
\ee
 At  $f>10^{11}$Hz the  vacuum term $|v_{k}|^2  \propto k^{-1}, k^{-3}, k^{-5}$ dominates,
the cross term
\be\label{alphbeta}
 Re [\alpha_k \beta_k^* v_k ^{2}] \propto  k^{-3}, k^{-5} ,
\ee
and
\be\label{betasqrp}
         |\beta_k|^2 |v_k |^2     \propto k^{-5}  ,
\ee
so that
\be\label{pdiv}
        p_k \propto k^4, k^2, k^0.
\ee
At  $f <10^{11}$Hz, the gravitons terms dominate.
These are illustrated in Fig. \ref{pressure2}.
Notice that,  by (\ref{rhohk}) and (\ref{pressk})  holding for the whole range $f>10^{-18}$Hz,
there is a relation
\be\label{relprho}
p_k(\tau) \simeq  \frac{1}{3} \rho_k(\tau),
\ee
i.e, $t^\mu\, _\mu(\tau) =0$.
Thus, $\rho_k $ and $p_k $ have the similar shape,
as seen in Figs. \ref{energy2} and \ref{pressure2}.

\section{Adiabatic Regularization of Divergences of Vacuum }

\subsection{power  spectrum  }

In  field theories,
divergences can occur in the expectation values
of physical quantities ,
such as the correlation function $\langle 0| h^{ij} h_{ij} |0 \rangle$
in Eq.(\ref{defspectrum})
and the energy-momentum tensor  $t_{\mu\nu}$ in Eq.(\ref{tmunu}) of RGW.
In Minkowski spacetime,
UV divergences of the vacuum can be removed by the normal ordering of the field operators.
However,   in curved spacetimes, the normal ordering does apply,
and the adiabatic regularization
 \cite{ParkerFulling1974,ParkerFullingHu1974,Bunch1980,AndersonParker1987, ParkerToms,BirrellDavies}
suits for removing  UV divergences of the vacuum.
Since the equation of RGW mode  $h_k(\tau)$
 has the same form as that of a minimally coupling,  massless scalar field,
and the regularization for the scalar field
\cite{ParkerToms,ParkerFulling1974,AndersonParker1987}
can be directly applied to RGW here.
From  (\ref{uvspectr}) we know that
the power spectrum has respective  quadratic and logarithmic  divergences.
By the minimal subtraction rule \cite{ParkerFulling1974,ParkerToms},
only  these two divergent parts are to be removed from the spectrum,
and the 2-nd adiabatic order subtraction is sufficient,
and one should not use the 4-th adiabatic order as claimed in Ref.\cite{Finelli2007}.
On the other hand, from  (\ref{rhodiv}) (\ref{pdiv}),
the energy density and pressure
contain  quartic divergences
besides the quadratic and logarithmic ones,
so that one should use the 4-th adiabatic order
as required by the minimal subtraction rule \cite{ParkerFulling1974,ParkerToms}.

We remark that
UV divergences  should not be simply dropped out as asserted in Ref.\cite{LythLiddle2009}.
Moreover, since RGW is regarded as a quantum field,
one can not  remove  UV divergences
by applying some smoothing technique, such as a window function,
which is often used for classical, stochastic fields.

There is an issue of  infrared divergence.
The   spectrum  of  (\ref{gspectrum})  in low frequency limit $k \rightarrow 0$
behaves as  $\Delta^2_t   \propto k^{2\beta+4}$,
and will also lead to the infrared divergence  in  the  correlation function.
The adiabatic regularization
has been developed, aiming at removing the UV divergence of vacuum,
not the IR divergence.
We shall not discuss the  issue in this paper.
See Refs \cite{FordParker1977,glenz,Allen1988} for further discussions.

To an  adiabatic $n$-th order,
the mode as a solution to Eq.(\ref{equ})
can be formally written as a general WKB function
\cite{ParkerToms,ParkerFulling1974,AndersonParker1987}
\be \label{wkb}
u^{(n)}_k(\tau)     =\frac{1}{\sqrt{2W^{(n)}(\tau)}}
            \exp\left[-i\int_{\tau_0}^{\tau}W^{(n)}(\tau')d\tau'\right],
\ee
where   $W^{(n)}(\tau)$ is a function for the adiabatic $n$th order.
For the massless minimally coupled scalar field \cite{AndersonParker1987},
       the 0-th order $W_k^{(0)}=k$,
and the 2-nd order
 \be \label{w2}
W_k^{(2)}=\sqrt{k^2-\frac{a''}{a}},
\ee
and the $n$-th adiabatic order\cite{Anderson2010}
\be \label{wn}
W_k^{(n)}=\sqrt{k^2-\frac{a''}{a}
-\frac{1}{2}\left[\frac{W_k^{(n-2)\prime\prime}}{W_k^{(n-2)}}
-\frac{3}{2}\left(\frac{W_k^{(n-2)\prime}}{W_k^{(n-2)}}\right)^2\right]}.
\ee
These formulae apply to RGW also.
The regularized spectrum is given by  \cite{Parker2007}
\be \label{respec}
\Delta_t^2(k,\tau)_{re}= A ^2\frac{k^3}{\pi^2 a^2}
   (|u_k(\tau)|^2-|u_k^{(2)}(\tau)|^2),
\ee
where
\be \label{counter}
|u_k^{(2)}|^2= \frac{1}{2W_k^{(2)}}
=\frac{1}{2k}\left(1-\frac{a''}{a k^2}\right)^{-\frac{1}{2}}
   \simeq \frac{1}{2k}+\frac{a''/a}{4k^3}
\ee
is the  counter term to the 2nd adiabatic order.
In principle, the regularization formula  (\ref{respec})
can apply at any time  $\tau$ during expansion.
When one chooses different time  $\tau$,
(\ref{respec}) will give different schemes of regularization.
Later in Sec.5 we shall consider three schemes.

As the first example, we apply regularization to the inflation stage.
In high-frequency limit $k\rightarrow\infty$,
the  mode $u_k(\tau)$ in  Eq.(\ref{u}) is expanded as
\ba \label{ukexpand}
u_k(\tau)&=&\frac{e^{-ik\tau}}{\sqrt{2k}}
(1-i\frac{\beta(\beta+1)}{2k\tau}-\frac{(\beta+2)(\beta+1)\beta(\beta-1)}{8k^2\tau^2}\nn\\
&&+i\frac{(\beta+3)(\beta+2)(\beta+1)\beta(\beta-1)(\beta-2)}{48k^3\tau^3}\nn\\
&&+\frac{(\beta+4)(\beta+3)(\beta+2)(\beta+1)\beta(\beta-1)(\beta-2)(\beta-3)}{384k^4\tau^4})
          +\mathcal{O}(k^{-\frac{11}{2}}), \nn\\
\ea
so that
\be \label{expu}
|u_k(\tau)|^2
        =\frac{1}{ 2k}
\left(1 + \frac{\beta(\beta+1)}{2k^2\tau^2}
      +\frac{3(\beta+2)(\beta+1)\beta(\beta-1)}{8 k^4 \tau^4} \right)
      +\mathcal{O}(k^{-7 }).
\ee
The first term $1/2k$ is quadratic divergence,
corresponding to the usual vacuum fluctuations in Minkowski space.
The second term is logarithmic divergence due to additional¡¡
vacuum fluctuations¡¡in expanding spacetime,
which can be written as
\be\label{logdiv}
\frac{\beta(\beta+1)}{ 4k^2\tau^2} = \frac{R}{24  (k/a)^2},
\ee
with the scalar curvature $R=6a''/a^3$.
This form  agrees with the known result
in the $R$-summed  and the  normal coordinate momentum space methods
       \cite{DeWitt1965,BunchParker1979,Parker1979b,ParkerToms}.
We   point out that the logarithmic divergence  (\ref{logdiv})
should not be written as a form $ (a'/a)^2/k^2$ of (1.5) in Ref.\cite{Agullo10}.
These two divergent terms are exactly canceled
by the 2-nd order adiabatic counter terms
$|u_k^{(2)} |^2    = \frac{1}{2k}+\frac{(\beta+1)\beta}{4k^3\tau^2}$
of  Eq.(\ref{counter}),
giving
\be \label{regvacuum}
|u_{k}|^2 - |u_k^{(2)}|^2
   = \frac{3(\beta-1)\beta (\beta+1) (\beta+2)}{16k^5  \tau^4}+\mathcal{O}(k^{-7}),
\ee
which comes from the  third  term in (\ref{expu}).
Thus, the  resulting  adiabatically-regularized spectrum
 in high frequencies  $f>10^{11}$Hz is
\be \label{Deltare}
\Delta_t^2(k,\tau)_{re} = A^2
\frac{3 (\beta-1)\beta (\beta+1) (\beta+2)}{16\pi^2 a^2 k^2\tau^4}
 \propto k^{-2} .
\ee
We plot  (\ref{Deltare}) for  $f> 10^{11}$Hz
in the top part of Fig. \ref{3rdregularizedspectrum}
          for the regularization at the end of inflation $\tau_1$.
As for low frequencies $f< 10^{11}$Hz,
the spectrum is less affected by the regularization.
When  (\ref{Deltare})  is substituted into (\ref{defspectrum}),
it gives a finite contribution to  the auto-correlation function
from the upper limit of integration.
(\ref{Deltare}) represents the  vacuum fluctuations at high frequencies
and has definite  physical effects.
As we shall see  in Sec 5,  (\ref{Deltare})
will evolve into the  high frequency  portion ($f>10^{11}$Hz) of the present spectrum,
and will serve as the target of high-frequency GW detectors \cite{Li2003,TongZhangGaussian}.

We like to clarify  two points regarding  adiabatic regularization.
First,
as it stands,  the subtraction by the counter terms in  (\ref{counter})
applies to the whole frequency range,
in contrary to what Ref. \cite{Durrer} suggested only for high frequencies,
though its effect on the spectrum is strong for
the high $k$-modes and weak for the low $k$-modes.
Second, by the the minimal subtraction rule,
the above 2-nd order regularization is sufficient for the power spectrum  \cite{Parker2007}.
If one tries to do a 4-th order adiabatic regularization
of the  spectrum   with   the factor
$\left(|u_k |^2-|u_k^{(2)}|^2\right)$ in  (\ref{respec})
replaced by     $   \left(|u_k |^2-|u_k^{(4)}|^2\right)$,
where the counterterm to the 4-th order is defined  in (\ref{4thorderterm}),
the regularized spectrum  would be infrared
divergent as   $ \propto  k^{- 2} $ as $k \rightarrow 0$.
This is  unacceptable.
Our calculation confirms the the minimal subtraction rule,
and  a 4-th order of adiabatic regularization is incorrect for the power spectrum.
This  conclusion is   just opposite to the  claim of  Ref.\cite{Finelli}.

In the special case of   de Sitter inflation with  $\beta=-2$,
the  analytical mode  (\ref{u}) is
\be \label{desol}
u_k(\tau)=\frac{e^{-ik\tau}}{\sqrt{2k}}(1-\frac{i}{k\tau}),
\ee
and $|u_k(\tau)|^2= 1/2k+ 1/2k^3\tau^2$,
which is just equal to the counter term $|u_k^{(2)}(\tau)|^2$,
resulting in a vanishing regularized spectrum  $\Delta_T^2(k)_{re}=0$.
This feature of regularization for  de Sitter
has been pointed out by Parker \cite{Parker2007} for a  massless scalar field.
However,
we shall show in Section 5
that  regularization   at present time
will save the spectrum  for the  de Sitter.

The next example is  for the accelerating stage,
in which RGW consists of both vacuum and gravitons.
Here consider only the vacuum,
whose mode is  (\ref{uvacpres}).
The calculations are similar to those the inflation stage.
One just replaces  $(1+\beta)$ by  $-\gamma$,
and  $\tau$ by $(\tau-\tau_a)$,
in (\ref{Deltare}),
arriving
the regularized   vacuum spectrum in high frequencies  $k\rightarrow\infty$
\be
\Delta_{vac}^2(k,\tau)_{re}
 = A^2   \frac{3(\gamma+1) \gamma (\gamma -1)(\gamma-2)}{16\pi^2 a^2 k^2  (\tau-\tau_a)^4}
 \propto k^{-2} .
\ee
The divergences of graviton part of the spectrum will addressed in  Section 5.1.

\subsection{  Vacuum Energy Density and Pressure of RGW}

 The vacuum  energy density and pressure of RGW
contain quartic, quadratic, and logarithmic divergences
as in (\ref{rhodiv}) (\ref{pdiv}),
which can be removed by the adiabatic regularization
to the   4th order \cite{ParkerFulling1974}.
For the spectral energy density, one takes
\be\label{rhoreg}
 \rho_k(\tau)_{re}=2\frac{k^3}{2\pi^2a^2}
             \left( | ( \frac{u_k(\tau)}{a} ) '  |^2
             -  | ( \frac{u_k^{(4)}(\tau) }{a} ) ' |^2  \right),
\ee
where the   counter  term   of the 4-th adiabatic order is
\be \label{adiabterms}
 | ( \frac{u_k^{(4)}(\tau) }{a} ) ' |^2=   \frac{|u_k^{(4)\prime}|^2}{a^2}
  + (\frac{a'}{a^2})^2 |u_k^{(4)}|^2
   -\frac{ a'}{a^3} ( u_k^{(4)*} u_k^{(4)\prime} + u_k^{(4)} u_k^{(4)*\prime} ) ,
\ee
and the 4th order adiabatic mode is
\be \label{wkb4}
u_k^{(4)}(\tau)= \frac{1}{\sqrt{2W_k^{(4)}(\tau)}}
          \exp\left[-i\int_{\tau_0}^{\tau}W_k^{(4)}(\tau')d\tau'\right]   .
\ee
Using the formula (\ref{wn}), one   calculates
(in Refs. \cite{Chakr,ParkerFulling1974,Bunch1980} $W_k^{(4)}$
      was computed for a massive scalar field),
\be\label{W4}
(W_k^{(4)})^2=k^2-\frac{a''}{a}-\frac{1}{4k^2a^2}
(a''^2-aa''''+2a'a'''-2\frac{a'^2a''}{a} ) ,
\ee
from which follow the  terms in (\ref{adiabterms}),
\ba \label{u4'}
\frac{1}{a^2}|u_k^{(4)\prime}|^2&=&\frac{W_k^{(4)}}{2a^2}
+\frac{(W_k^{(4)\prime})^2}{8(W_k^{(4)})^3a^2}\nn\\
  &&\simeq \frac{k}{2a^2}-\frac{a''/a}{4a^2k}
  -\frac{1}{16k^3a^4}(2a''^2-aa''''+2a'a'''-2\frac{a'^2a''}{a})
      \nn\\
\ea
\be \label{a'2counter}
(\frac{a'}{a^2})^2|u_k^{(4)}|^2 = (\frac{a'}{a^2})^2 \frac{1}{2W_k^{(4)}}
\simeq (\frac{a'}{a^2})^2\left(\frac{1}{2k}+\frac{a''/a}{4k^3}\right)    ,
\ee
\be\label{a'counter}
 -\frac{ a'}{a^3}(u_k^{(4)*} u_k^{(4)\prime} + u_k^{(4)} u_k^{(4)*\prime})
  \simeq\frac{1}{4k^3}
 \left(\frac{a'^2a''}{a^5}-\frac{a'''a'}{a^4}\right)   ,
\ee
and the counter term is
\be\label{countu4}
  | ( \frac{u_k^{(4)}(\tau) }{a} ) ' |^2 =  \frac{k}{2a^2}
 +\frac{1}{4a^2k}\left(\frac{2a'^2}{a^2}-\frac{a''}{a}\right)
 +\frac{1}{8a^2k^3}\left(\frac{5a'^2a''}{a^3}-\frac{a''^2}{a^2}+\frac{a''''}{2a}
 -\frac{3a'a'''}{a^2}\right)  .
\ee
Substituting these into (\ref{rhoreg}) gives the regularized spectral energy density.
In a similar fashion for the spectral pressure, one takes
\be\label{pressure}
p_k(\tau)_{re}= \frac{k^5}{3\pi^2a^4}  \left(|u_k(\tau)|^2-|u_k^{(4)}|^2\right),
\ee
where
\be
|u_k^{(4)}|^2=\frac{1}{2W_k^{(4)}}
  \simeq \frac{1}{2k}+\frac{a''/a}{4k^3}+\frac{1}{16k^5a^2}
   (   4a''^2-aa''''+2a'a'''- \frac{2a'^2a''}{a}  )
\ee
is the 4th order adiabatic counter term.

The above regularization formulae hold for any time $\tau$.
First apply  to the inflation stage,
during which
the energy density and pressure of RGW
have only the vacuum contributions.
Using  the mode $u_k$ of  (\ref{u}),
  one has
\ba
 | ( \frac{u_k(\tau)}{a} ) '  |^2
 &=&
 \frac{1}{a^2}\left[\frac{k}{2}+\frac{(\beta+1)(\beta+2)}{4k\tau^2}
 +\frac{3\beta(\beta+1)(\beta+2)(\beta+3)}{16k^3\tau^4}\right. \nonumber  \\
&& \left.+\frac{5(\beta-1)\beta(\beta+1)(\beta+2)(\beta+3)(\beta+4)}{32k^5\tau^6}\right]
               +O(k^{-7}), \nonumber
\ea
and,  by (\ref{countu4}),  the counter term is
\[
 | ( \frac{u_k^{(4)}(\tau) }{a} ) ' |^2
   =\frac{1}{a^2}\left[\frac{k}{2}+\frac{(\beta+1)(\beta+2)}{4k\tau^2}
 +\frac{3\beta(\beta+1)(\beta+2)(\beta+3)}{16k^3\tau^4}\right],
\]
which just cancel
the quartic, quadratic, and logarithmic divergences of
$ \left|\left( u_k(\tau)/ a \right) ' \right|^2$,
yielding  the regularized spectral energy density in high-frequency limit
\ba \label{reenergy}
 \rho_k(\tau)_{re}=
 \frac{5(\beta-1)\beta(\beta+1)(\beta+2)
               (\beta+3)(\beta+4)}{32 \pi^2 a^4  k^2\tau^6} +O(k^{-4}).
\ea
For the pressure, similar calculations give
\ba
 \left|  u_k(\tau)   \right|^2
 & = &   \frac{1}{2 k}  + \frac{\beta(\beta+1) }{4 k^3 \tau^2}
    + \frac{3(\beta -1 ) \beta (\beta+1) (\beta+2)}{16 k^5 \tau^4} \nonumber \\
&&  + \frac{ 5(\beta-2)(\beta-1)\beta (\beta  +1)(\beta+2)(\beta+3) }{32 k^7\tau^6}
    +O(k^{-8}),    \nonumber
\ea
and
\be\label{4thorderterm}
 \left| u_k^{(4)}(\tau)    \right|^2  = \frac{1}{2 k}  + \frac{\beta(\beta+1) }{4 k^3 \tau^2}
    + \frac{3(\beta -1 ) \beta (\beta+1) (\beta+2)}{16 k^5 \tau^4} ,
\ee
and the regularized spectral pressure  in high-frequency limit
\be    \label{rp}
p_k(\tau)_{re}
  = \frac{5(\beta-2)(\beta-1)\beta(\beta+1)(\beta+2)(\beta+3)}{3 \times 32\pi^2 a^4 k^2 \tau^6}
   +O(k^{-4}).
\ee
The expressions (\ref{reenergy}) (\ref{rp})
hold only at  $f > 10^{11}$Hz.
The low frequency  $f < 10^{11}$Hz parts of $\rho_k(\tau)_{re}$ and $p_k(\tau)_{re}$
are less affected by adiabatic regularization.
(\ref{reenergy}) and (\ref{rp})
tell that  $\rho _{k\, re}$, $p_{k\, re}$    $\propto k^{-2}$
 at  high frequencies,
and   the relation (\ref{relprho}) is modified to
\be
 p_k(\tau)_{re} =  \frac{1}{3}\frac{\beta-2}{\beta+4}\rho_k(\tau)_{re},
\ee
i.e, there is a trace anomaly of regularized  RGW,
\be
t^\mu\, _{\mu}(\tau)_{re} =\frac{6}{(\beta+4)}\rho_k(\tau)_{re} \ne 0
\ee
at high frequencies.
This situation of anomaly is similar to what happens to
a conformally-coupling massless scalar field after regularization
\cite{ParkerToms,BirrellDavies,BLHU1978,Ford1985}.
Notice that (\ref{reenergy}) (\ref{rp}) give
  $ \rho _{k\, re} >0$ and    $ p_{k\, re} <0$ for $\beta < -2$,
which is similar to the inflaton  field that drives inflation.
However, the magnitude of vacuum fluctuations are very small,
$ \rho _{k\, re} /\rho \sim (H /M_{Pl})^2 \sim 10^{-16}$,
where $\rho$ be the inflation energy scale with  $\rho^{1/4} \sim 10^{15}$Gev.
In passing, we notice that
a negative pressure of quantum fields at infrared ranges
 also arises in the context discussed  in Ref.\cite{FordParker1977}.

Next apply to the accelerating stage.
Consider the  vacuum part of the energy density and pressure,
i.e, the $v_k(\tau)$  parts of (\ref{u'2}) and (\ref{p}).
Just replacing  $(\beta +1) \rightarrow -\gamma $ in (\ref{reenergy}) and (\ref{rp}),
one obtains  the regularized energy density and pressure
of the present vacuum at high frequencies $f > 10^{11}$Hz
\ba \label{vacreenergy}
 \rho_k(\tau)_{vre}=
 \frac{5(\gamma-3)(\gamma-2)(\gamma-1)\gamma(\gamma+1)(\gamma+2)}{32 \pi^2 a^4  k^{2} (\tau-\tau_a)^6}
    +O(k^{-4}) ,
\ea
\be \label{vacprr}
p_k(\tau)_{vre}
  = \frac{5(\gamma-2)(\gamma-1)\gamma(\gamma+1)(\gamma+2)(\gamma+3)}
    {3 \times 32\pi^2 a^4 k^{2} (\tau-\tau_a)^6}.
   +O(k^{-4}),
\ee
The accelerating model can be fitted by $\gamma\simeq 2.1$,
(\ref{vacreenergy}) and (\ref{vacprr})
give $\rho _{k\, re}< 0 $ and $ p _{k\, re} >0$ for   $\gamma >2$.
However, as will be seen in the next section,
(\ref{vacreenergy}) and (\ref{vacprr}) from the vacuum
are  overwhelmed by those of gravitons,
so that the total spectral energy density and pressure of RGW
are positive at high frequencies.

\

\section{Regularization at different time }

As defined  in Eq.(\ref{spectrum}),
the spectrum $\Delta_t(k,\tau)$ depends on time.
At what time  $\tau$ should  regularization be performed ?
In literature, there have been disagreements
 in the context
 of a scalar inflaton \cite{Agullo09,Agullo10, Durrer},
and the issue also exists for  RGW.
In the following, we shall   regularize in three methods, respectively:
at the time of observation,
 at the end of inflation \cite{Durrer},
and    at horizon crossing time for each $k$-mode  \cite{Agullo09,Agullo10}.
As it turns out,
the three resulting regularized spectra are quite
similar  in regard to observations.

\subsection{Regularization at the present time}

\begin{figure}
\centering
\includegraphics[width=0.7\linewidth]{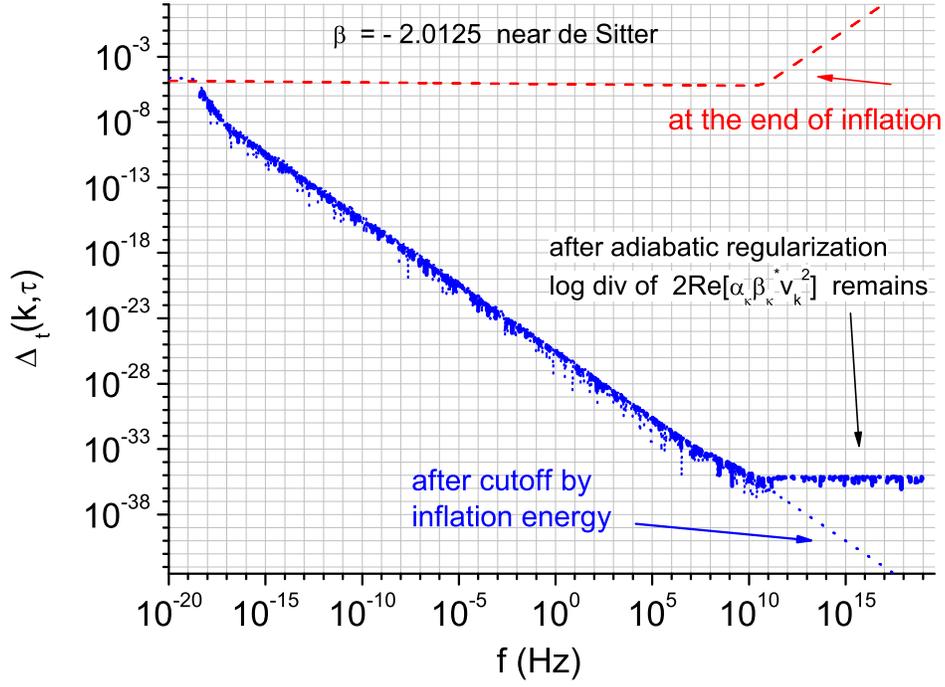}
\caption{Regularization at the present time $\tau_H$.
The unregularized spectrum at the end of inflation is at the top,
and the present spectrum  is at the lower part.
}
 \label{spectrumatdifftimes}
\end{figure}

\begin{figure}
\centering
\includegraphics[width=0.7\linewidth]{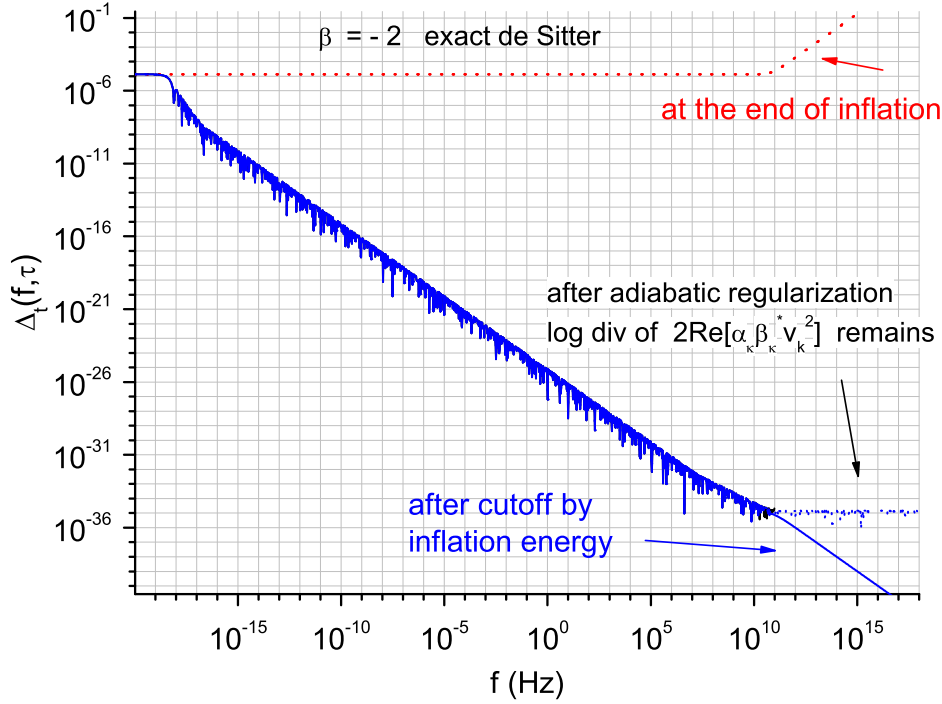}
\caption{
The spectrum regularized at present is  non-vanishing for $\beta=-2$,
which would be zero if regularized during inflation  \cite{Parker2007}.
}
 \label{spectrumatdifftimes2}
\end{figure}

This is the first scheme of regularization.
Suppose an observer is to   detect RGW now,
and it is natural  to regularize the  spectrum
at the present time $\tau_H$ of observation during the accelerating stage.
To do this,
one  just sets   $\tau=\tau_H$
in the expression (\ref{respec}) of regularized spectrum
\be \label{regularpresent}
\Delta_t^2(k,\tau_H)_{re} =A^2  \frac{k^3}{\pi^2 a(\tau_H)^2}
   (|u_k(\tau_H)|^2-|u_k^{(2)}(\tau_H)|^2),
\ee
where the mode function
 $u_k(\tau_H)$ is  explicitly given by  Eq.(\ref{upresent}),
and the counter term $|u_k^{(2)}(\tau_H)|^2$ by (\ref{counter}).
Using   (\ref{uuvac}) at $\tau_H$,   one has
\be \label{reguk}
|u_k|^2- |u_k^{(2)}|^2
= (|v_{k}|^2- |u_k^{(2)}|^2 )   + 2Re[\alpha_k \beta_k^*  v_k^2]
       + 2|\beta_k|^2  |v_k|^2      ,
\ee
where $(|v_{k}|^2- |u_k^{(2)}|^2 ) \propto k^{-5}$ is the regularized vacuum part,
already known   in (\ref{regvacuum}),
and the graviton part
is unaffected by the adiabatic regularization.
The  regularized   spectrum $\Delta_t (k,\tau_H)_{re}$ of (\ref{regularpresent})
is plotted  in Fig.\ref{spectrumatdifftimes} for   $\beta=-2.0125$.
The non-vanishing spectrum of de Sitter case  $\beta=-2$
is plotted  in Fig.\ref{spectrumatdifftimes2},
which, as mentioned earlier,  would be vanishing if the regularization
is  performed during inflation \cite{Parker2007}.

After adiabatic regularization,
the  logarithmic divergence   due to gravitons
 of $2Re[\alpha_k \beta_k^*  v_k^2]$ term in (\ref{reguk}) still remains,
as is indicated by a flat curve at $f>10^{11}$Hz
in Figs. \ref{spectrumatdifftimes} and \ref{spectrumatdifftimes2}.
{
 It is well known that  the occurrence of this kind of UV divergence is caused by
 the discontinuity of $a''(\tau)$ between two adjacent stages of the model,
 thus can be removed by choices of continuous $a''(\tau)$.
 L. Ford  [8] demonstrated this by an  explicit example,
in which  a finite time duration $\Delta t$  of the transition is assumed,
and some  smooth $a(\tau)$ with continuous $a''$ is constructed.
These resulted in a graviton number density
 $ \propto \ln(\frac{1}{H\Delta t})$,  which is finite.
In our model the discontinuity of $a''$
corresponds to an abrupt    transition with $\Delta t=0$,
so there is no surprise that
 UV divergences appear in the spectrum,
   graviton number density,  energy density, pressure, etc.}
To remove this artificial  divergence,
 Ref.\cite{Allen1988}  proposed that
gravitons  are not produced with higher energy
than the  inflation energy scale, say $\sim 10^{16}$GeV.
This yields a   cutoff of
the  logarithmic divergence  of gravitons at $f> 10^{11}$Hz.
Here we adopt this simple treatment.
Thus, after adiabatic regularization and  cutoff as well,
the spectrum  becomes convergent,
 $\Delta^2 _t(k,\tau_H)_{re} \propto k^{-2}$  for $f> 10^{11}$Hz
plotted as a dotted line in the lower left part
in Fig.\ref{spectrumatdifftimes} and Fig.\ref{spectrumatdifftimes2}, respectively.

For $f<10^{11}$Hz,
 the spectrum is contributed by gravitons,
 and essentially unchanged by regularization and cutoff.
In particular,
the primordial spectrum defined at the low frequency end $f< 10^{-18}$Hz
remains   the same as  (\ref{powerlaw}).
In a manner similar to (\ref{index}) and  (\ref{running}),
one can also define  the regularized spectral indices
\be
n_{t\, re}  \equiv
          \frac{d\ln\Delta_t^2(k)_{re}}{d\ln k}| _{k \rightarrow 0}= 2\beta+4
\ee
and
\be
\alpha_{t\, re} \equiv
  \frac{d^2 \ln \Delta^2_t(k)_{re}}{d(\ln k)^2}|_{k \rightarrow 0} =0,
\ee
both have the same values as those in (\ref{index}) and (\ref{running}), respectively.

\begin{figure}
\centering
\includegraphics[width=0.7\linewidth]{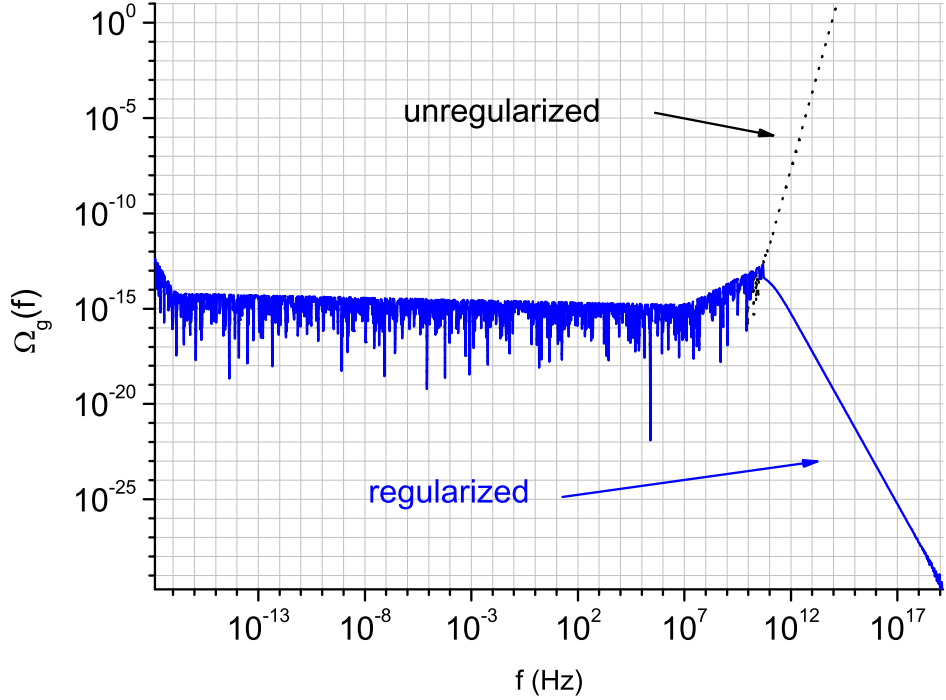}
\caption{Regularized spectral energy density
          $\Omega_g (k)_{re} = \rho_k(\tau_H)_{re}/\rho_c$  at  present. }
\label{enspectrum}
\end{figure}

\begin{figure}
\centering
\includegraphics[width=0.7\linewidth]{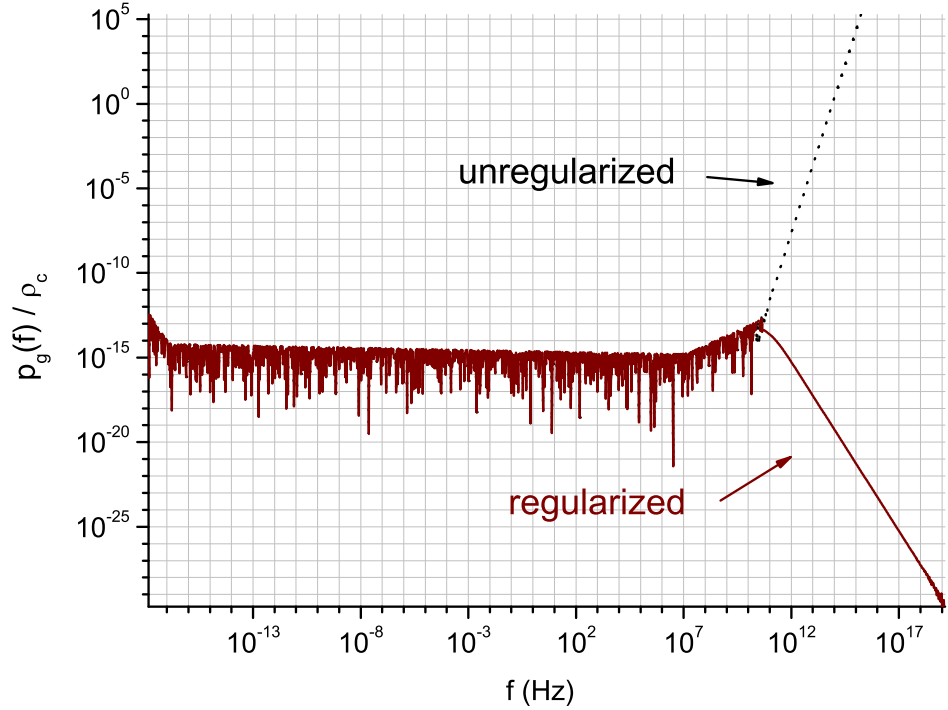}
\caption{Regularized spectral pressure $p_k(\tau_H)_{ re}/ \rho_c$  at  present.   }
   \label{press}
\end{figure}

Now the energy density and pressure at present.
The spectral energy density
is adiabatically regularized to the 4-th order by
\ba \label{om}
 \rho_k(\tau_H)_{re}=\frac{k^3}{\pi^2a^2}
         \left(  | ( \frac{v_k }{a} ) '  |^2
             -  | ( \frac{u_k^{(4)}  }{a} ) ' |^2
    +  2Re [\alpha_k \beta_k^*   ( \frac{v_k }{a} ) ^{\prime\,  2} ]
    +  2|\beta_k|^2   | ( \frac{v_k }{a} ) '  |^2 \right),
       \nonumber \\
\ea
where $|\left( \frac{v_k }{a}\right)'|^2 - |(\frac{u_k^{(4)}}{a}) ' |^2$
is the regularized vacuum part,
known in (\ref{vacreenergy}) at high frequencies,
and the regularized spectral pressure is
\be \label{prpres}
p_k(\tau_H)_{re}  = \frac{k^5}{3\pi^2a^4}\left( (|v_{k}|^2- |u_k^{(4)}|^2 )
    + 2Re( \alpha_k \beta_k^*  v_k ^2 )   + 2|\beta_k|^2  |v_k   |^2  \right),
\ee
where   $(|v_{k}|^2- |u_k^{(4)}|^2 )$
is known in  (\ref{vacprr})
at high frequencies.
The divergences of $ \rho_k$ and $p_k$ due to gravitons have been known in
  (\ref{rhogravitona}),  (\ref{rhogravitonb}), (\ref{alphbeta}), (\ref{betasqrp}),
which, unaffected by regularization,
will be cutoff  for $f>10^{11}$Hz  by the same argument of the inflation energy
as for the power spectrum.
After    cutoff, the graviton part of
$ \rho _{k\, re}$ and  $p _{k\, re} $ at  $ f>10^{11}$Hz
are given by the following leading terms,
\ba\label{cutoffenergy}
 \rho _{k\, gr} =3 p _{k\, gr}
& \simeq & \frac{k^4}{\pi^2a^4}  \left[\frac{\beta(\beta+1)}{4x_1^2}
\left(\frac{(\beta-1)(\beta+2)}{4x_1}+\frac{\beta_s(\beta_s+1)}{4t_1}\right)\right.\nn\\
&-&\left.\frac{\beta_s(\beta_s+1)}{4t_1^2}
\left(\frac{\beta(\beta+1)}{4x_1}+\frac{(\beta_s-1)(\beta_s+2)}{4t_1}\right)\right]^2
                     \propto k^{-2}  ,   \nn\\
\ea
where $x_1 \equiv k |\tau_1| $ and $t_1 \equiv k|\tau_1-\tau_p|$.
In comparison, this  is  many orders   higher than
the  present vacuum energy   (\ref{vacreenergy}).
Thus,
the total spectral energy  after regularization and cutoff
is  positive for $f>10^{11}$Hz.
Over  the broad range  $(10^{-18}-10^{11})$Hz,
$ \rho _{k\, re}  , p _{k\, re} $ are dominated by   gravitons,
and remain practically  unchanged after regularization.
The final  $ \rho _{k\, re}$, $p _{k\, re} $ after regularization and cutoff
are plotted in Figs.\ref{enspectrum} and   \ref{press}, respectively.

\subsection{Regularization at the end of inflation}

Next, we explore the second scheme of regularization,
in which
the spectrum of RGW is regularized at the time $\tau_1$, the end of inflation  \cite{Durrer}.
Moreover, we shall also let the associated, regularized mode $u^{re}_k(\tau)$
evolve subsequently according to its field  equation,
and arrive at the present spectrum at the time $\tau_H$.
Since the regularized spectrum vanishes for exact de Sitter inflation,
we consider the general case $\beta\ne -2$.
We apply  the general formula (\ref{respec}) of regularized spectrum
at  the end of inflation with  $\tau=\tau_1$,
\be \label{regattau1}
|u_k^{re}(\tau_1)|^2  = |u_k(\tau_1)|^2 -|u^{(2)}_k(\tau_1)|^2,
\ee
which has been known in  Eq.(\ref{Deltare}).
This fixes the amplitude of  regularized mode $u_k^{re}(\tau_1) $  at $\tau=\tau_1$
as the initial condition.
To determine its phase,
assume that $u_k^{re}(\tau_1)$ has the same
phase as the unregularized mode  $u_k(\tau_1)$,
which has the following asymptotic behavior
\ba \label{asymptotic}
&&u_k(\tau_1)\simeq \frac{e^{-ik\tau_1}}{\sqrt{2k}},
\ \ \ \ \ \ \text{for}\ k\rightarrow\infty,  \nn\\
&&u_k(\tau_1)\simeq
 \frac{\sqrt{-\pi\tau_1}e^{i(\beta-1)\pi/2}}{2\Gamma(\beta+\frac{3}{2})\cos\beta\pi }
           \left(\frac{k\tau_1}{2}\right)^{\beta+\frac{1}{2}}
      +O( k^{ \beta +\frac{3}{2} } ) \ \  \ \ \ \text{for}\ k\rightarrow 0.\nonumber \\
\ea
So we choose  the initial condition at $\tau_1$ to be
 \be \label{ukretau1}
u_k^{re}(\tau)=e^{i\theta}\sqrt{|u_k(\tau)
|^2-\frac{1}{2k}\left(1 + \frac{\beta(\beta+1)}{2k^2\tau^2}\right)},
 \,\,\,\,  \tau =\tau_1
\ee
where the phase $\theta$ is approximately chosen as
\be
\theta=\begin{cases} -k\tau \ \ \ \ \text{for}\ k>-\tau_1^{-1}
\\ (\beta-1)\pi/2 \ \ \ \ \text{for}\ k <-\tau_1^{-1} ,
          \end{cases}
\ee
and the time-derivative  is given by
$u_k^{re\, '}(\tau)= \frac{d}{d\tau} u_k^{re }(\tau)$ at $  \tau =\tau_1 $.
The modes $u_k(\tau)$  before $\tau_1$ during inflation
remain the same as that in  Eq.(\ref{u}).
The subsequent evolution of $u_k(\tau)$ follows straightforwardly,
from reheating, radiation, matter, to the accelerating stage.
As has been checked,
other choices of $\theta$
will give the same outcome of spectrum at present.
Although the phase of mode  contains
 information of quantum state of RGW,
in addition to the spectrum \cite{Grishchuk},
 the current detectors, such as LIGO,
are not capable of detecting such a phase information \cite{AllenFlaganan2000}.
The above treatment of the phase is sufficient for our purpose
since we  are mainly concerned  with the spectrum.

\begin{figure}
\centering
\includegraphics[width=0.7\linewidth]{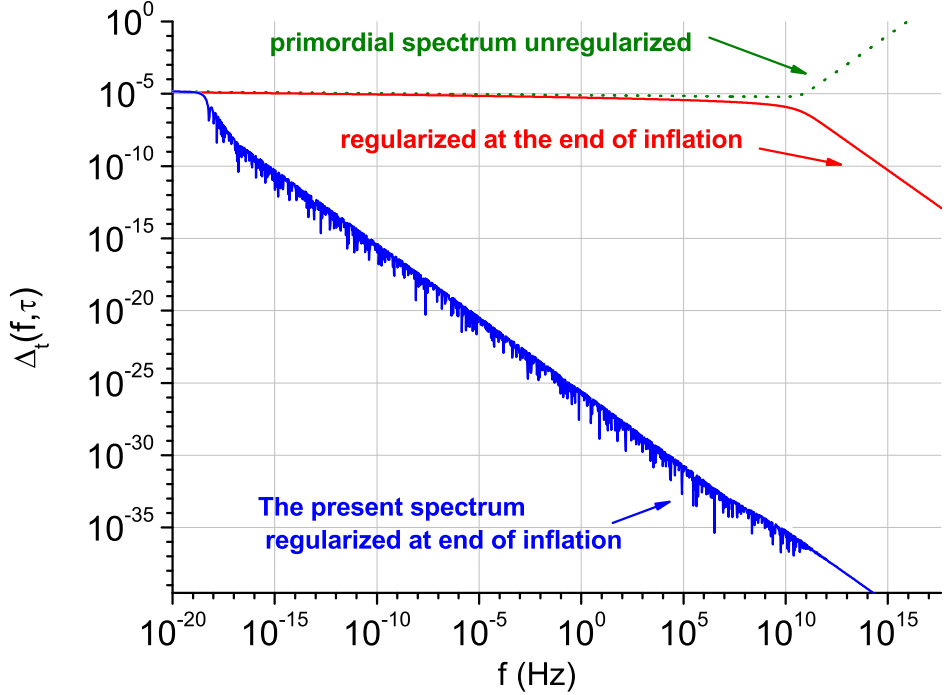}
\caption{The spectrum is regularized at the end of inflation,
then evolves into the present spectrum,
which turns out to be almost the same as that of Fig.\ref{spectrumatdifftimes}
 regularized at the present time.
  }
 \label{3rdregularizedspectrum}
\end{figure}

The regularized, initial  spectrum at the end of inflation
and the  present spectrum   evolved from the former
are plotted  in Fig.\ref{3rdregularizedspectrum}.
Notice that
the regularized spectrum at present
 behaves as   $\Delta^2_t(k)_{re}\propto f^{-2}$ at high frequencies $f> 10^{11}$Hz.
Thus, all the UV divergences of power spectrum have been all removed by
this scheme of adiabatic regularization,
including  the quadratic and logarithmic divergences of vacuum
and the  logarithmic divergence of gravitons as well.
No additional cutoff is needed,
in contrast to the first scheme in the last subsection 5.1.
Comparing with the first scheme,
 the amplitudes differ by about $20$ around $f\sim 10^{11}$Hz,
and  by about $1.2$ around $f\sim 10^{0}$Hz.
At the low frequency end $f\sim 10^{-18}$Hz,
 the primordial spectrum,
the indices $n_t$,  and $\alpha_t$
remain  unchanged, as in  (\ref{powerlaw}),  (\ref{index}) and (\ref{running}),
respectively.
The resulting spectra at present in these two schemes are
essentially  the same.

\subsection{Regularization at horizon exit }

Finally,  we explore the regularization at horizon-exit
proposed in Ref.\cite{Agullo09,Agullo10}.
Again we  consider the case $\beta\ne -2$.
The low-frequency modes of RGW exit  the horizon during inflation
at a time  $|\tau_k|  = 1/k$,
and regularization is performed at $\tau_k$ for each mode.
So in this scheme
the regularization time is not instantaneous but rather at different $\tau_k$
for different modes.
By  the formula of (\ref{respec}) during inflation,  one obtains directly
\be \label{respec2}
\Delta_{t}^2(k)_{re}=C(\beta)\frac{k^{2\beta+4}}{\pi^2M_{Pl}^2l_0^2}
 \ \ \ \ \text{for}\ k< \frac{1}{|\tau_1| } ,
\ee
with
$C(\beta)\equiv\pi\left|H^{(2)}_{\frac{1}{2}+\beta}(1)\right|^2-(2+\beta(\beta+1))$.
This can be rewritten as
\be \label{respec3}
\Delta_{t}^2(k)_{re}=C^2 \Delta_{t}^2(k) \ \ \ \ \text{for}\ k< \frac{1}{| \tau_1| } ,
\ee
where  $\Delta_{t}^2(k)$ is the primordial spectrum in (\ref{powerlaw})
and
\be \label{reunre}
  C^2 \equiv 2^{2\beta+1} \pi^{-1} \Gamma(\beta+\frac{3}{2})^2\cos^2(\beta\pi)  C(\beta)
  \simeq 0.904 \cdot |2+\beta|.
\ee
depending on $\beta $, and  $C ^2\simeq0.01$ for $\beta=-2.0125$.
For the modes that have exited the horizon,
the subsequent evolution of  the regularized modes is given by
\be\label{eta}
u^{re}_k(\tau)=C  u_k(\tau)\ \ \ \ \text{for}\ k< \frac{1}{ |\tau_1| }.
\ee

\begin{figure}
\centering
\includegraphics[width=0.7\linewidth]{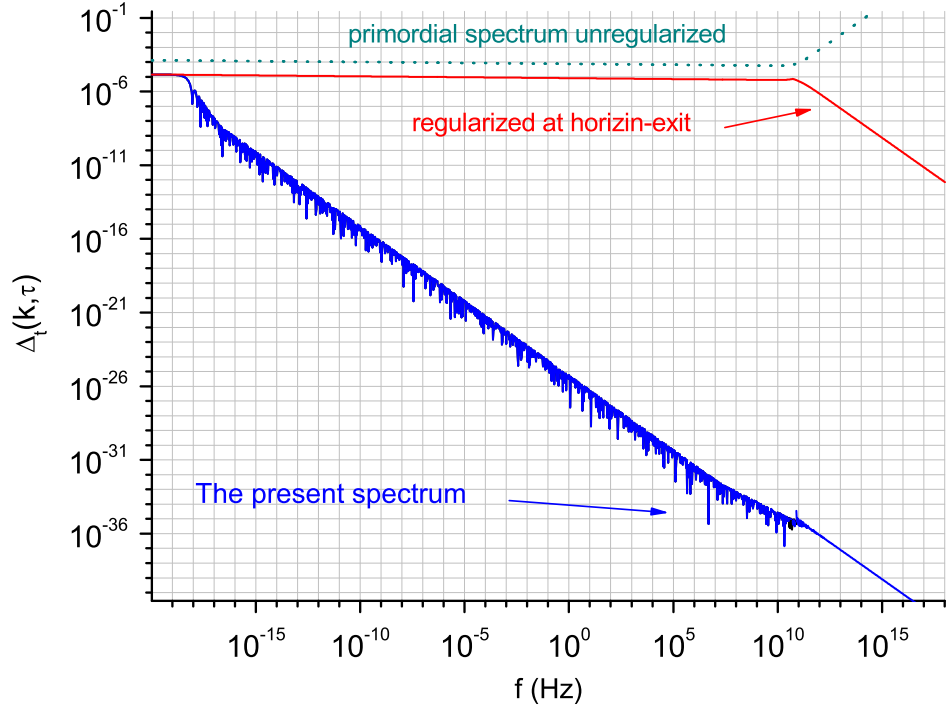}
\caption{The spectrum is regularized at horizon-exit,
then evolves into the present spectrum,
which is similar to that of Fig.\ref{spectrumatdifftimes} and Fig.\ref{3rdregularizedspectrum},
but with a lower, overall amplitude.
  }
 \label{spectrumtoday2}
\end{figure}

On the other hand,
the high-frequency modes with  $k> 1/|\tau_1|$
never exit  the horizon during inflation.
It is these modes that give rise to the UV divergence  of the spectrum.
Ref.\cite{Agullo09,Agullo10} did not give a treatment of these  modes.
We   regularize these  modes at the end of inflation $ \tau=\tau_1$,
\be \label{highmode}
u_k^{re}(\tau)  =e^{-ik\tau }\sqrt{|u_k(\tau)|^2-\frac{1}{2k}
\left(1 + \frac{\beta(\beta+1)}{2k^2\tau^2}\right)}\ \
             \ \ \ \text{for}\ k> 1/|\tau_1|,
\ee
and the time-derivative of the modes  is given by
$u_k^{re\, '}(\tau)= \frac{d}{d\tau} u_k^{re }(\tau)$ at $  \tau =\tau_1 $.
Thus, the primordial spectrum regularized in the third scheme is
\be \label{spectrumatt1}
\Delta_t^2(\tau_1,k)_{re}=\begin{cases}
    2 A^2 \frac{k^3}{2\pi^2 a^2}| C{u_k(\tau_1)}|^2\ \ \ \ \text{for}\ k< 1/|\tau_1|
\\2 A^2 \frac{k^3}{2\pi^2 a^2}\left[|u_k(\tau_1)|^2-|u_k^{(2)}(\tau_1)|^2\right]\ \ \ \
         \text{for}\ k>  1/|\tau_1|   \end{cases},
\ee
as shown in the upper part of Fig.\ref{spectrumtoday2}.
Given the initial condition   (\ref{highmode}),
the subsequent evolution of the  high-frequency modes $u_k(\tau)$ is also determined.
The  regularized,  present spectrum $\Delta_{t}^2(k,\tau_H )_{re}$
is shown in the lower part of Fig.\ref{spectrumtoday2}.
The profile is similar to those in the first and second schemes,
but with the amplitude lower by the factor $C^2$.
At low-frequency end,
\be\label{DeltaC}
\Delta_{t}^2(k,\tau_H)_{re}=C^2 \Delta_{t}^2(k)  \ \ \ {\rm for}  \ k<   1/ |\tau_1|
\ee
where $\Delta_{t}^2(k) $ is the  primordial spectrum in (\ref{powerlaw}),
and $\Delta_{t}^2(k)  \propto H^2 \propto \rho$,
where $\rho$ is the energy density  of inflation.
Since the  amplitude will be eventually fixed by CMB observations as  in (\ref{initial}),
 one just  raises $\rho$ of the   model by a factor $1/C^2$
to achieve the same amplitude as in the first and second schemes.

Although (\ref{DeltaC})  formally resembles
the  result  of slow-roll approximation in Ref.\cite{Agullo09},
nevertheless, $C^2$ is independent of $k$ here.
Therefore,
 the spectral indices $n_t$ and   $\alpha_t$ are
unaffected by this regularization
and remain the same as   in Eqs.(\ref{index}) and (\ref{running}).
We emphasize that no  slow-roll approximation  is involved in  our result.

\section{ Regularization of Primordial Spectra of Inflaton and Scalar Curvature Perturbation}

The above result of RGW and regularization for the inflation stage
can be extended to the scalar metric perturbation
via an  inflaton field.
The scale factor $a (\tau) \propto |\tau|^{1+\beta}$ in (\ref{inflation})
has been taken as a generic inflation model so far.
It can be specifically realized by a single scalar inflaton  field,
also called the power-law inflation
                  \cite{Lucchin1985,AbbottWise1984,LythStewart1992,StewartLyth1993}.
In this case the exact solution of perturbed scalar field is the same form as RGW.
For inflation   driven by a scalar field $\phi$,
the  Friedmann equations are
\be \label{Friedmann1}
\phi''+2\mathcal{H}\phi'+a^2\frac{\partial V}{\partial \phi}=0,
\ee
\be \label{Friedmann2}
\mathcal{H}^2=\frac{8\pi G}{3}\left(\frac{1}{2}\phi'^2+a^2V\right),
\ee
where  $\mathcal{H}=\frac{a'}{a}$ and $V$ is the potential.
For $a (\tau) \propto |\tau|^{1+\beta}$,
the solution of Eqs.(\ref{Friedmann1}) and (\ref{Friedmann2})
is
\[
\frac{\phi}{M_{Pl}}=\sqrt{2(1+\beta)(2+\beta)}\ln(-\tau) +B ,
\]
\be
V(\phi)=V_0\exp\left(-\sqrt{\frac{2\beta+4}{\beta+1}}\frac{\phi}{M_{Pl}}\right).
\ee
The slow-roll parameters are defined by
\be \label{epsleta}
\epsilon\equiv\frac{M_{Pl}^2}{2}\left(\frac{V'(\phi)}{V}\right)^2=\frac{\beta+2}{\beta+1}
\ \ \ \ \ \ \eta\equiv M_{Pl}^2\frac{V''(\phi)}{V}=2\epsilon.
\ee
Consider the perturbed scalar field $\delta\phi$,
which generally depends on the choice of coordinates.
In fact, from $\delta\phi $ and scalar metric perturbations,
one can construct a gauge-invariant perturbed scalar field $\bar{\delta\phi}$,
satisfying the  field equation
              \cite{Hwang1994,Hwang1996,Gordon2000}
\be \label{final7}
\bar{\delta\phi} '' +2\mathcal{H}\bar{\delta\phi} '
-\nabla^2 \bar{\delta\phi}+\left[a^2\frac{\partial^2V}{\partial\phi^2}
       -\frac{8\pi G}{a^2}\frac{d}{d\tau}
\left(\frac{a^2\phi'\,^2}{\mathcal H }\right)\right] \bar{\delta\phi}=0,
\ee
which  holds in any coordinates.
For  $a (\tau) \propto |\tau|^{1+\beta}$,
  the bracket in Eq.(\ref{final7})  vanishes,
so that  the equation reduces to
\be \label{fieldeq}
w_k''+(k^2-\frac{a''}{a})w_k=0,
\ee
 where  $  \bar{\delta\phi} _k \equiv w_k/ a$ for each $k$-mode.
 Eq.(\ref{fieldeq})  is identical to Eq.(\ref{equ}) of RGW,
and its solution $w_k$ is the same as   $u_k$ in  (\ref{u}).
Thus,  we obtain the spectrum  of $\bar{\delta\phi}$
\be
\Delta^2_{\bar{\delta\phi}} (k,\tau)
   =\frac{k^3}{2\pi^2}|\bar{\delta\phi} _k|^2
   = \frac{k^3}{2\pi^2 a^2 }|w_k|^2
   =\frac{M^2 _{Pl}}{8} \Delta^2_t(k,\tau),
\ee
where $ \Delta^2_t(k,\tau) $ is the RGW spectrum in (\ref{gspectrum}).
The spectral indices are given
$n_{\bar{\delta\phi}} -1
  \equiv \frac{d \ln \Delta^2_{\bar{\delta\phi}}(k)}{d\ln k}|_{k \rightarrow 0}
                          =2\beta+4$ ,
$\alpha_{\bar{\delta\phi}}
  \equiv  \frac{d^2 \ln \Delta^2_{\bar{\delta\phi}} (k)}{d(\ln k) ^2}|_{k \rightarrow 0}
   =0$.
One is more interested in scalar metric perturbations,
which are  directly related to observations.
Introducing the curvature perturbation
 $R \equiv \frac{{\mathcal H } }{\phi\, '}\bar{\delta\phi} $,
which is also gauge invariant  \cite{KodamaSasaki1984,Sasaki1986,Mukahanov1988,Mukahanov2005},
the  spectrum of scalar curvature perturbation   is given by
\be \label{scalarspec}
\Delta^2_R(k,\tau)=\frac{k^3}{2\pi^2}|R_k|^2
=\frac{k^3}{2\pi^2a^2}\frac{\mathcal H^2}{\phi^{'2}} |w_k|^2
=  \frac{1+\beta}{8(2\beta+4)} \Delta^2_t(k,\tau) ,
\ee
where $\frac{{\mathcal H } }{\phi\, '}  =\frac{1}{M_{Pl}}\sqrt{\frac{1+\beta}{2\beta+4}}$
is a constant in our model.
The primordial spectrum of $R$
relevant to CMB observations is defined at far  outside horizon,
\be \label{srscalar}
\Delta_R ^2(k) \equiv  \Delta^2_R(k,\tau)   |_{k\ll 1/|\tau|}
       =\frac{1}{2 \epsilon M_{Pl}^2}a_t^2 \left(\frac{H}{2\pi}\right)^2 k^{2\beta+4}.
\ee
The scalar spectral index is given  by
\be \label{sindex}
n_s-1\equiv \frac{d \ln \Delta^2_R(k)}{d\ln k}|_{k \rightarrow 0}
=2\beta+4 =\frac{-6\epsilon}{1-\epsilon}+\frac{2\eta}{1-\eta/2}
\ee
 differing from the result of slow-roll approximation   $n_s  = 1 -6\epsilon+2\eta$
in Ref.\cite{LiddleLythbook},
and the scalar running index is
\be\label{runreg}
\alpha_s \equiv  \frac{d^2 \ln \Delta^2_R(k)}{d(\ln k) ^2}|_{k \rightarrow 0}
   =0,
\ee
  differing  from the result
  $\alpha_s =-16\epsilon \eta +24\epsilon^2+2\xi^2$
with $\xi^2 \equiv  M_{pl}^4 \frac{V' V'''}{V^2}$
in slow-roll approximation in Ref.\cite{KosowskyTurner}.
The  tensor-scalar-ratio  is
\be \label{ratio}
r   \equiv  \frac{\Delta^2_t(k)}{\Delta^2_R(k)}=16\frac{\beta+2}{\beta+1} =16\epsilon,
\ee
and the consistency relation is
\be\label{consistrel}
r =\frac{-8n_t}{1-n_t/2} .
\ee
Both (\ref{ratio}) (\ref{consistrel}) are valid for the whole relevant range of $k$.
The adiabatic regularization in (\ref{respec}) for RGW
is directly adopted here,
yielding the regularized scalar curvature spectrum,
\be \label{PR}
\Delta^2_R(k,\tau )_{re}= \frac{1+\beta}{2\beta+4} \frac{k^3}{2\pi^2a^2M_{Pl}^2}
\left(|w_k|^2-\frac{1}{2k}-\frac{a''/a}{4k^3}\right).
\ee
For  regularization at the end of inflation,
similar to Section 5.2 for RGW,
the primordial scalar spectrum, $n_s$, $\alpha_s$  and $r$ all remain unchanged.
For regularization at horizon-exit,
the spectrum becomes
\be \label{PR3}
\Delta^2_R(k)_{re}=C^2 \Delta^2_R(k) \,\,\, {\rm for}\,\, k\ll 1/|\tau_1|,
\ee
with  $C^2 $ given by Eq.(\ref{reunre}),
thus  $n_s$, $\alpha_s$, $r$  and  remain unchanged.
These results  differ  from that of
the slow-roll approximation in Ref.\cite{Agullo09,Agullo10}.

\section{Conclusion and Discussion }

Three aspects of RGW have been studied in this paper:
the analytic spectrum and spectral indices,
the   decomposition of the present RGW  into vacuum and gravitons,
and the removal of UV divergences of the spectrum, energy density, and pressure,
arising from both vacuum and gravitons.
Similarly,
regularization  during inflation
is also performed for the gauge-invariant  perturbed inflaton
and the scalar curvature perturbation.

The analytical  $\Delta^2_t(k,\tau)$,  $n_t$,  $\alpha_t$, and  $r$
that have been obtained
are  valid  at any time and frequencies during inflation.
The  exact relations involving the indices in (\ref{index3}) (\ref{sindex})  (\ref{consistrel})
differ from  those in the slow-roll approximation,
and can  be  tested by future CMB observations.
In particular,
using the observed spectra $C^{XX}_l$ of CMB in   $l\simeq (10-3000)$,
$\Delta^2_t(k,\tau)$ is actually constrained at $ k|\tau_1| \sim (10^{-28} -10^{-26})$,
 far outside the horizon during inflation.
This is where the primordial spectrum (\ref{powerlaw}) is referred.
The spectrum  at the horizon exit $k|\tau|=1$
 differs drastically from the primordial one.

The present RGW as quantum field in curved spacetime
actually consists of the vacuum and graviton parts.
This decomposition sheds light on
the structure of spectrum  over the respective frequency ranges.
The present vacuum
gives quadratic, logarithmic UV divergences to $\Delta^2_t(k,\tau_H)$,
and the gravitons   give a logarithmic divergence,
at high-frequency end  $f>10^{11}$Hz.
However,  at $f<10^{11}$Hz
 gravitons are  dominant,
and the current detectors are all operating within this band.
In this sense,
these detectors are to detect gravitons,  not the present vacuum of RGW.
The  graviton number density  $|\beta_k|^2 $
is  contributed by all four discontinuity points of $a''(\tau)$,
among which the inflation-reheating transition is the greatest.

In removing  UV divergences of RGW,
we have carried out regularization in three schemes:
at the present time,
at the end of inflation,
and at horizon-exit during inflation.
The first scheme actually involves two parts.
The adiabatic regularization removes only
the divergences of present vacuum,
and the divergences of gravitons are  cutoff.
The last two schemes remove the vacuum divergences during inflation,
the regularized spectrum is then taken as the initial condition
and evolves into  the present   spectrum,
which is convergent.
Besides,
for the spectrum, the 2-nd order adiabatic regularization is sufficient
to remove UV divergences,
and the 4-th adiabatic order is not used
according to  the minimal subtraction rule.
In all these  three schemes,
the regularized, present  spectra are similar,
 except that the third scheme yields a lower  amplitude,
which can be raised by a higher inflation energy scale
in  confronting observational data.
At high frequencies $f>10^{11}$Hz,
the  three regularized spectra behave as  $\Delta^2_t(k,\tau_H)_{re} \propto  k^{-2}$,
which can serve as a target for high-frequency GW detectors,
{
such as  the polarized, Gaussian laser beam detectors
proposed in  Refs.\cite{Li2003, TongZhangGaussian}.
}
At     $f< 10^{11}$Hz,
the regularized  spectrum  remains practically unchanged.

We also calculate the spectral energy density  and  pressure  of RGW.
The vacuum  $\rho_k$ and $p_k$
contain  quartic, quadratic and logarithmic divergences,
and the regularization to 4-th adiabatic order is necessary and sufficient
to remove them.
The regularized vacuum $\rho_{k,\,  re} > 0$ and  $p_{k\, re} < 0$
at high frequencies   for inflation with $\beta < -2$,
which is implied by the current observations.
For the present accelerating stage with  $\gamma\simeq 2.1$,
 the vacuum $\rho_{k,\,  re}< 0$ and $p_{k\, re}>0$.
The graviton part of  $\rho_k$ and $p_k$ at present
contain only quadratic and logarithmic divergences,
which are cut off by the inflation energy scale,
yielding  $ \rho _{k\, gr} =3 p _{k\, gr} \propto k^{-2}$  at $f>10^{11}$Hz,
 greater than those of vacuum by many orders.
At    $f< 10^{11}$Hz,
$\rho_k$ and $p_k$ are practically unchanged by regularization and cutoff.
Hence the  total spectral energy density and pressure, after regularization and cutoff,
 are dominantly contributed by gravitons over the whole frequency range.

{
Now we give our assessment of the three schemes of regularization.
During the course of cosmic expansion,
 the actual, physical spectrum is  described by
the  instantaneously-regularized  $( |u_k(\tau)|^2- |u_k^{(2)}(\tau)|^2 )$
 for any instance $\tau$.
This instantaneous regularization
is closest to the instantaneous normal-ordering for quantum field in flat spacetime.
  The two schemes, at the present time  or at the end of inflation,
are different  demonstrations of this same physical quantity
 at respective instances.
Moreover,  as our work  has shown,
their  resulting two spectra for observation at present are nearly the same.
So we think  that   these two schemes
are  better choice
than the one at horizon exit,
since the latter  is not instantaneous but at different times for different $k$-modes.
Among the two instantaneous schemes,
we would like to remark that,
in regard to  technical convenience,
 the first scheme, i.e. regularization at the present time
 for   observation,
   is a natural choice
  since it is simpler than the one at the end of inflation.  }

For the scalar field driving the power-law inflation,
the gauge invariant perturbed scalar field
has an exact solution which is the same as RGW,
so does the scalar curvature perturbation.
The regularization is the same as for RGW,
in particular, the scalar spectral indices $n_s$ and $\alpha_s$,
tensor-scalar ratio $r$ remain unchanged.

 Our study has an advantage that
it is based on the exact solutions of RGW for the whole expansion,
and of the   perturbed inflaton
and the scalar curvature perturbation during inflation.
Our  results  have demonstrated,
Parker-Fuling's adiabatic regularization and the minimal subtraction rule
work  perfectly well in removing  UV divergences of vacuum,
to the 2-nd order for the spectrum,
to  the 4-th order for the energy density and pressure, respectively.

\

\textbf{Acknowledgements}

Y. Zhang is supported by NSFC Grant No. 11275187, NSFC 11421303,
SRFDP, and CAS,
the Strategic Priority Research Program
``The Emergence of Cosmological Structures"
of the Chinese Academy of Sciences, Grant No. XDB09000000.

\

\

\appendix
\numberwithin{equation}{section}

{\bf \large Appendix}

\

In this appendix,
we list the explicit expressions of mode functions of RGW
and the associated coefficients,
 which are determined analytically
from continuous joining of $h_k(\tau)$ and $h'_k(\tau)$
for all five consequent stages of cosmic expansion.

The scale factor  $a(\tau)$ in a power-law form is given  for all five stages,
so that both $a(\tau)$and its time derivative $a\,'(\tau)$
are continuous  at the four transition points between all five stages.
{ 
But, $a''$ is not required to be continuous for simplicity.
 This  simple modeling of $a(\tau)$ 
 works well for  our purpose of an exact RGW solution,
but the artificial discontinuity of $a''$ would bring about
 too much graviton production, as addressed in Sec.5.1.
 }

There is an overall normalization of $a(\tau)$.
In this paper, we take $|\tau_H-\tau_a|=1$.
The present Hubble radius $H^{-1}_0= l_H/\gamma= 9.257\times 10^{28}h^{-1}$cm
with the Hubble parameter $h =0.69$.
This fixes the parameters in the expressions of $a(\tau)$.
Furthermore,  for the time of the transitions,
we use  the following cosmological specifications in our computation:
The matter-accelerating transition time  at a redshift $z\sim 0.347 $.
The radiation-matter transition taken at $z =3293$ \cite{Komatsu}.
The inflation energy scale is taken to $10^{16}$Gev,
the reheating duration is chosen such that  $a(\tau_s)/a(\tau_1)= 300$,
so that at the beginning of radiation the energy scale will be $10^{13}$Gev.
Details of the specifications have been given  in Refs. \cite{zhangyang05,Zhang06}.

\

\

The inflation stage:
The scalae factor $a(\tau)$ is in Eq.(\ref{inflation}),
  the mode of RGW is in Eq.(\ref{u})  as part of the initial condition.

The reheating stage:
\be
a(\tau)=a_z|\tau-\tau_p|^{1+\beta_s},\,\,\,\,\tau_1\leq \tau\leq \tau_s.
\ee
As a model parameter, we  simply take $\beta_s= -0.3$.
The general solution of Eq.(\ref{evolution}) during the reheating  stage is
\be \label{reheatingu}
u_k(\tau ) = \sqrt{\frac{\pi}{2}}\sqrt{\frac{t}{2k}}
     \big[ b_1(k) H^{(1)}_{ \beta_s+ \frac{1}{2} } (t)
          +b_2(k) H^{(2)}_{\beta_s+ \frac{1}{2}} ( t) \big],
         \,\,\,\,\, \, \tau_1 <\tau\leq \tau_s,
\ee
where $t=k(\tau-\tau_p)$.
The two coefficients $b_1$ and $b_2$ are determined by the joining condition at $\tau_1$
\ba
b_1(k)&=\Delta_b^{-1}\left\{ \sqrt{\frac{x_1}{t_1}}
\left[a_1H^{(1)}_{\beta+\frac{1}{2}}(x_1) +a_2H^{(2)}_{\beta+\frac{1}{2}}(x_1)\right]
\left[\frac{1}{2\sqrt{t_1}}H^{(2)}_{\beta_s + \frac{1}{2}}(t_1)+
\sqrt{t_1} H^{(2)\prime}_{\beta_s +\frac{1}{2}}(t_1)\right] \right. \nn\\
&\left.-H^{(2)}_{\beta_s + \frac{1}{2}}(t_1)
\left[\frac{1}{2\sqrt{x_1}}\left(a_1H^{(1)}_{\beta+\frac{1}{2}}(x_1)
+a_2H^{(2)}_{\beta+\frac{1}{2}}(x_1)\right)\right.\right.\nn\\
&\left.\left.+\sqrt{x_1}\left(a_1H^{(1)\prime}_{\beta+\frac{1}{2}}(x_1)
+a_2H^{(2)\prime}_{\beta+\frac{1}{2}}(x_1)\right) \right] \right\},
\ea
\ba
b_2(k)&=\Delta_b^{-1}\left\{ \sqrt{\frac{x_1}{t_1}}
\left[a_1H^{(1)}_{\beta+\frac{1}{2}}(x_1) +a_2H^{(2)}_{\beta+\frac{1}{2}}(x_1)\right]
\left[\frac{1}{2\sqrt{t_1}}H^{(1)}_{\beta_s + \frac{1}{2}}(t_1)+
\sqrt{t_1} H^{(1)\prime}_{\beta_s +\frac{1}{2}}(t_1)\right] \right. \nn\\
&\left.-H^{(1)}_{\beta_s + \frac{1}{2}}(t_1)
\left[\frac{1}{2\sqrt{x_1}}\left(a_1H^{(1)}_{\beta+\frac{1}{2}}(x_1)
+a_2H^{(2)}_{\beta+\frac{1}{2}}(x_1)\right)\right.\right.\nn\\
&\left.\left.+\sqrt{x_1}\left(a_1H^{(1)\prime}_{\beta+\frac{1}{2}}(x_1)
+a_2H^{(2)\prime}_{\beta+\frac{1}{2}}(x_1)\right) \right] \right\},
\ea
\[
\Delta_b= \sqrt{t_1}\left[H^{(1)}_{\beta_s+ \frac{1}{2} } (t_1) H^{(2)'}_{ \beta_s+ \frac{1}{2}   } (t_1)
       -H^{(1)'}_{ \beta_s+ \frac{1}{2} } (t_1) H^{(2)}_{ \beta_s+ \frac{1}{2} } (t_1)\right].
\]
where $x_1=k\tau_1$ and $t_1=k(\tau_1-\tau_p)$,
while the coefficients $a_1$ and $a_2$ are given by Eq.(\ref{a1a2}).
In the high frequency limit $k\rightarrow\infty$
\ba
b_1(k)&=&i\left(\frac{\beta(\beta+1)}{4x_1^2}-\frac{\beta_s(\beta_s+1)}{4t_1^2}\right)
e^{-i(x_1+t_1)+i\pi\beta+i\pi\beta_s/2}+\mathcal{O}(k^{-3})
\ea
\ba
b_2(k)&=&-ie^{-i(x_1-t_1)+i\pi\beta-i\pi\beta_s/2}
\left(1-i\frac{\beta(\beta+1)}{2x_1}+i\frac{\beta_s(\beta_s+1)}{2t_1}
-\frac{\beta^2(\beta+1)^2}{8x_1^2}\right.\nonumber\\
&&\left.-\frac{\beta_s^2(\beta_s+1)^2}{8t_1^2}+\frac{\beta(\beta+1)\beta_s(\beta_s+1)}{4x_1t_1}
\right) +\mathcal{O}(k^{-3})
\ea

The radiation-dominant stage:
\be \label{ra}
a(\tau)=a_e(\tau-\tau_e),\,\,\,\,\tau_s\leq \tau\leq \tau_2.
\ee
and the mode function is
\be
u_k(\tau ) = \sqrt{\frac{\pi}{2}}\sqrt{\frac{y}{2k}}
     \big[ c_1(k) H^{(1)}_{\frac{1}{2} } (y)
          +c_2(k) H^{(2)}_{\frac{1}{2} } ( y) \big],
         \,\,\,\,\, \, \tau_s <\tau\leq \tau_2,
\ee
where $y=k(\tau-\tau_e)$ and $c_1$ and $c_2$ are given by
\ba
c_1(k)&=\Delta_c^{-1}\left\{ \sqrt{\frac{t_s}{y_s}}
\left[b_1H^{(1)}_{\beta_s+\frac{1}{2}}(t_s) +a_2H^{(2)}_{\beta_s+\frac{1}{2}}(t_s)\right]
\left[\frac{1}{2\sqrt{y_s}}H^{(2)}_{\frac{1}{2}}(y_s)+
\sqrt{y_s} H^{(2)\prime}_{\frac{1}{2}}(y_s)\right] \right. \nn\\
&\left.-H^{(2)}_{ \frac{1}{2}}(y_s)
\left[\frac{1}{2\sqrt{t_s}}\left(a_1H^{(1)}_{\beta_s+\frac{1}{2}}(t_s)
+a_2H^{(2)}_{\beta_s+\frac{1}{2}}(t_s)\right)\right.\right.\nn\\
&\left.\left.+\sqrt{t_s}\left(a_1H^{(1)\prime}_{\beta_s+\frac{1}{2}}(t_s)
+a_2H^{(2)\prime}_{\beta_s+\frac{1}{2}}(t_s)\right) \right] \right\},
\ea
\ba
c_2(k)&=\Delta_b^{-1}\left\{ \sqrt{\frac{t_s}{y_s}}
\left[a_1H^{(1)}_{\beta_s+\frac{1}{2}}(t_s) +a_2H^{(2)}_{\beta_s+\frac{1}{2}}(t_s)\right]
\left[\frac{1}{2\sqrt{y_s}}H^{(1)}_{\frac{1}{2}}(y_s)+
\sqrt{y_s} H^{(1)\prime}_{\frac{1}{2}}(y_s)\right] \right. \nn\\
&\left.-H^{(1)}_{\frac{1}{2}}(y_s)
\left[\frac{1}{2\sqrt{t_s}}\left(a_1H^{(1)}_{\beta_s+\frac{1}{2}}(t_s)
+a_2H^{(2)}_{\beta_s+\frac{1}{2}}(t_s)\right)\right.\right.\nn\\
&\left.\left.+\sqrt{t_s}\left(a_1H^{(1)\prime}_{\beta_s+\frac{1}{2}}(t_s)
+a_2H^{(2)\prime}_{\beta_s+\frac{1}{2}}(t_s)\right) \right] \right\},
\ea
\[
\Delta_c= \sqrt{y_s}\left[H^{(1)}_{\frac{1}{2} } (y_s) H^{(2)'}_{\frac{1}{2}} (y_s)
       -H^{(1)'}_{\frac{1}{2} } (y_s) H^{(2)}_{\frac{1}{2} } (y_s)\right].
\]
where $t_s=k(\tau_s-\tau_p)$ and $y_s=k(\tau_s-\tau_e)$.
In the high frequency limit $k\rightarrow\infty$
\ba
c_1(k)&=&i\left(\frac{\beta(\beta+1)}{4x_1^2}-\frac{\beta_s(\beta_s+1)}{4t_1^2}\right)
e^{-i(x_1+t_1-t_s+y_s)+i\pi\beta}\nonumber\\
&&+i\frac{\beta_s(\beta_s+1)}{4t_s^2}e^{-i(x_1-t_1+t_s+y_s)+i\pi\beta}+\mathcal{O}(k^{-3}),
\ea
\ba
c_2(k)&=&-ie^{-i(x_1-t_1+t_s-y_s)+i\pi\beta}
\left(1-i\frac{\beta(\beta+1)}{2x_1}+i\frac{\beta_s(\beta_s+1)}{2t_1}
-i\frac{\beta_s(\beta_s+1)}{2t_s}\right.\nonumber\\
&&\left.-\frac{\beta^2(\beta+1)^2}{8x_1^2}-\frac{\beta_s^2(\beta_s+1)^2}{8t_1^2}-\frac{\beta_s^2(\beta_s+1)^2}{8t_s^2}
+\frac{\beta(\beta+1)\beta_s(\beta_s+1)}{4x_1t_1}\right.\nonumber\\
&&\left.-\frac{\beta(\beta+1)\beta_s(\beta_s+1)}{4x_1t_s}
+\frac{\beta_s^2(\beta_s+1)^2}{4t_1t_s}
\right) +\mathcal{O}(k^{-3}).
\ea

The matter-dominant stage:
\be \label{m}
a(\tau)=a_m(\tau-\tau_m)^2,\,\,\,\,\tau_2 \leq \tau\leq \tau_E.
\ee
and the mode function is
\be
u_k(\tau ) = \sqrt{\frac{\pi}{2}}\sqrt{\frac{z}{2k}}
     \big[d_1(k) H^{(1)}_{\frac{3}{2} } (z)
          +d_2(k) H^{(2)}_{\frac{3}{2} } (z) \big],
         \,\,\,\,\, \, \tau_2 <\tau\leq \tau_E,
\ee
where $z=k(\tau-\tau_m)$ and $d_1$ and $d_2$ are given by
\ba \label{d-1}
d_1(k)&=\Delta_d^{-1}\left\{ \sqrt{\frac{y_2}{z_2}}
\left[b_1H^{(1)}_{\frac{1}{2}}(y_2) +a_2H^{(2)}_{\frac{1}{2}}(y_2)\right]
\left[\frac{1}{2\sqrt{y_s}}H^{(2)}_{\frac{3}{2}}(z_2)+
\sqrt{z_2} H^{(2)\prime}_{\frac{3}{2}}(z_2)\right] \right. \nn\\
&\left.-H^{(2)}_{ \frac{3}{2}}(z_2)
\left[\frac{1}{2\sqrt{y_2}}\left(a_1H^{(1)}_{\frac{1}{2}}(y_2)
+a_2H^{(2)}_{\frac{1}{2}}(y_2)\right)\right.\right.\nn\\
&\left.\left.+\sqrt{y_2}\left(a_1H^{(1)\prime}_{\frac{1}{2}}(y_2)
+a_2H^{(2)\prime}_{\frac{1}{2}}(y_2)\right) \right] \right\},
\ea
\ba \label{d-2}
d_2(k)&=\Delta_d^{-1}\left\{ \sqrt{\frac{y_2}{z_2}}
\left[a_1H^{(1)}_{\frac{1}{2}}(y_2) +a_2H^{(2)}_{\frac{1}{2}}(y_2)\right]
\left[\frac{1}{2\sqrt{z_2}}H^{(1)}_{\frac{3}{2}}(y_s)+
\sqrt{z_2} H^{(1)\prime}_{\frac{3}{2}}(z_2)\right] \right. \nn\\
&\left.-H^{(1)}_{\frac{3}{2}}(z_2)
\left[\frac{1}{2\sqrt{y_2}}\left(a_1H^{(1)}_{\frac{1}{2}}(y_2)
+a_2H^{(2)}_{\frac{1}{2}}(y_2)\right)\right.\right.\nn\\
&\left.\left.+\sqrt{y_2}\left(a_1H^{(1)\prime}_{\frac{1}{2}}(y_2)
+a_2H^{(2)\prime}_{\frac{1}{2}}(y_2)\right) \right] \right\},
\ea
\[
\Delta_d= \sqrt{z_2}\left[H^{(1)}_{\frac{3}{2} } (z_2) H^{(2)'}_{\frac{3}{2}} (z_2)
       -H^{(1)'}_{\frac{3}{2} } (z_2) H^{(2)}_{\frac{3}{2} } (z_2)\right].
\]
where $z_2=k(\tau_2-\tau_m)$ and $y_2=k(\tau_2-\tau_e)$.
In the high frequency limit $k\rightarrow\infty$
\ba
d_1(k)&=&-\left(\frac{\beta(\beta+1)}{4x_1^2}-\frac{\beta_s(\beta_s+1)}{4t_1^2}\right)
e^{-i(x_1+t_1-t_s+y_s-y_2+z_2)+i\pi\beta}\nonumber\\
&&-\frac{\beta_s(\beta_s+1)}{4t_s^2}e^{-i(x_1-t_1+t_s+y_s-y_2+z_2)+i\pi\beta}\nonumber\\
&&+\frac{1}{2z_2^2}e^{-i(x_1-t_1+t_s-y_s+y_2+z_2)+i\pi\beta}+\mathcal{O}(k^{-3}),
\ea
\ba
d_2(k)&=&-e^{-i(x_1-t_1+t_s-y_s+y_2-z_2)+i\pi\beta}
\left(1-i\frac{\beta(\beta+1)}{2x_1}+i\frac{\beta_s(\beta_s+1)}{2t_1}
-i\frac{\beta_s(\beta_s+1)}{2t_s}\right.\nonumber\\
&&\left.+i\frac{1}{z_2}-\frac{\beta^2(\beta+1)^2}{8x_1^2}-\frac{\beta_s^2(\beta_s+1)^2}{8t_1^2}-\frac{\beta_s^2(\beta_s+1)^2}{8t_s^2}
-\frac{1}{2 z_2^2}+\frac{\beta(\beta+1)\beta_s(\beta_s+1)}{4x_1t_1}\right.\nonumber\\
&&\left.-\frac{\beta(\beta+1)\beta_s(\beta_s+1)}{4x_1t_s}
+\frac{\beta(\beta+1)}{2x_1z_2}+\frac{\beta_s^2(\beta_s+1)^2}{4t_1t_s}
-\frac{\beta_s(\beta_s+1)}{2t_1z_2}+\frac{\beta_s(\beta_s+1)}{2t_s z_2}
\right) \nonumber\\&&+\mathcal{O}(k^{-3}).
\ea

The accelerating stage up to the present time $\tau_H$:
\be \label{accel}
a(\tau)=l_H|\tau-\tau_a|^{-\gamma},\,\,\,\,\tau_E \leq \tau\leq \tau_H.
\ee
with   $\gamma \simeq 2.1$
fits the model $\Omega_\Lambda \simeq 0.7$ and $\Omega_m =1-\Omega_\Lambda $.
The  mode function $u_k(\tau)$ and the coefficients $\alpha_ k$, $\beta_ k$
of this stage
are given in (\ref{upresent}), (\ref{e-1}) and  (\ref{e-2}).

\

\end{document}